\documentclass[11pt]{article}

\usepackage{geometry}
\usepackage[table,xcdraw]{xcolor}
\usepackage{color}  

\usepackage{floatrow} 
\newfloatcommand{capbtabbox}{table}[][\FBwidth]

\usepackage{setspace}
\usepackage{caption} 
\usepackage{graphicx}
\usepackage{multirow}

\usepackage{arydshln} 

\definecolor{tred}{HTML}{de2d26}
\usepackage{colortbl}

\newcommand{\reddashedline}{\arrayrulecolor{black}\hdashline\arrayrulecolor{black}}
\usepackage{lscape} 

\usepackage{mathptmx}
\usepackage[T1]{fontenc}

\usepackage[natbibapa]{apacite}

\newenvironment{biseabstract}{%
\begin{quote} \bf}
{\end{quote}}

\newenvironment{bisekeywords}{%
\begin{quote} \it \textbf{Keywords -}}
{\end{quote}}

\title{Constructing Effective Customer Feedback Systems -- \\ A Design Science Study Leveraging Blockchain Technology} 

\author
{Mark C. Ballandies$^{1\ast}$, Valentin Holzwarth$^{2}$, Barry Sunderland$^{3}$, Evangelos Pournaras$^{4}$ and \\ Jan vom Brocke$^{2}$\\
\\
\normalsize{$^{1}$Computational Social Science, ETH Zurich,}\\
\normalsize{Stampfenbachplatz 50, Zurich, Switzerland}\\
\normalsize{$^{2}$Hilti Chair of Business Process Management, University of Liechtenstein, Liechtenstein}\\
\normalsize{$^{3}$ETH Library Lab, ETH Zurich}\\
\normalsize{$^{4}$School of Computing, University of Leeds}\\
\\
\normalsize{$^\ast$To whom correspondence should be addressed; E-mail:  bcmark@protonmail.com.}
}

\date{}

\begin{document} 

\baselineskip24pt


\maketitle

\begin{biseabstract} 
Organizations have to adjust to changes in the ecosystem, and customer feedback systems (CFS) provide important information to adapt products and services to changing customer preferences. 
However, current systems are limited to single-dimensional rating scales and are subject to self-selection biases. 
This work contributes design principles for CFS and implements a CFS that advances current systems by means of contextualized feedback according to specific organizational objectives. 
We apply Design Science Research (DSR) methodology and report on a longitudinal DSR journey considering multiple stakeholder values by utilizing value-sensitive design methods. We conducted expert interviews, design workshops, demonstrations, and a four-day experiment in an organizational setup, involving 132 customers of a major Swiss library. In the process, we validated the identified design principles and the implemented software artifact both qualitatively and quantitatively and drew conclusions for their efficient instantiation. 

In particular, we found that i) blockchain technology can afford three design principles of effective CFS. Also, ii) combining DSR with value-sensitive design methods explicitly provides rationale for design principles in the form of identified important values.  Moreover, iii) utilizing this methodology makes the construction of software artifacts more efficient it terms of design time by restricting the design space of a software artefact to those options that align with stakeholder values. 
Hence, the findings of this work advance the knowledge on the design of CFS and provides both, for researchers a theoretical contribution to reason about design principles and a guideline to managers and decision makers for designing software artefacts efficiently. 
  
\end{biseabstract}

\begin{bisekeywords}
Design Science Research; Blockchain; Feedback System; Cryptoeconomics; value-sensitive Design; Token Engineering
\end{bisekeywords}



\section{Introduction}
\label{sec:introduction}
Customer feedback is important for the potential of an organization to differentiate itself from competitors \citep{culnan1989designing} and to improve its products and services according to customer preferences \citep{stoica2015mining,hu2016factors}.
Nevertheless, due to status differences, hierarchy steepness, and reduced levels of cooperation, large hierarchical organizations such as firms or public institutions impede the flow of feedback to and within their organization \citep{anderson2010functions}, which reduces the quality of management decisions \citep{khatri2009consequences}.

Inspired by the observation that technology can be utilized to support the self-organization capacity and thus the efficiency of a large scale system via feedback loops \citep{ashby1964introduction}, as it has been shown for a traffic light control network \citep{lammer2008self}, this work generates design knowledge on the construction of a customer feedback system (CFS) that improves the provision of high-quality feedback about services and products from customers to an organization. 

We report on Design Science Research at a case organization, which is a major Swiss library. This library is challenged by a lack of feedback from so-called unaware-customers, i.e. customers that are not aware that they are using services provided by the organization. Moreover, the library is challenged to distinguish important from unimportant feedback, particularly, when the feedback quantity is high. Furthermore, the library has difficulties evaluating the questions utilized in solicited surveys with customers. The innovation \& networking team within the library organization had been mandated to implement a solution in the form of a CFS that improves the status quo of feedback provision from library users to the organization. We identified this need of the library in the first step of the applied research methodology (Section \ref{sec:methodology}) and consecutively accompanied the construction of the CFS. 

We argue that a CFS should not only optimize for performance, but its design also needs to integrate the values of stakeholders \citep{van2015handbook,kleineberg2021social} such as autonomy or credibility. This has been recognized by the IS community \citep{friedman2013value,maedche2017interview}. Though already utilized in similar systems \citep{friedman2002informed,miller2007value}, value considerations in the methods of CFS construction and thus the resulting design knowledge is limited. In particular, to our best knowledge, design principles for CFS that explicitly consider values have not been found. We therefore ask the first Research Question (RQ1):

\textit{(RQ1) What are the design principles of a value-sensitive customer feedback system?}

By applying an established design science research (DSR) methodology \citep{sonnenberg2011evaluation,sonnenberg2012evaluations,hevner2004design,peffers2007design} and putting a focus on stakeholder values during the design phase as performed in \citet{ballandies2021}, we facilitate both, i) the value-alignment of the created tool with the affected stakeholders and ii) the implementation of the design principles in a software artifact, which is iteratively evaluated at different stages. 
A controlled experiment with 132 customers of the library and focus groups with experts of that organization are conducted to measure the performance of the software artifact in terms of usability and quality of collected feedback. In order to evaluate this, we ask the second Research Question (RQ2):

\textit{RQ2: What is usability and quality of collected feedback of a software artifact that implements the design principles?}

This paper illustrates how design science as a journey can be conducted \citep{vom2020special} and contributes the following: 
I) design principles for a value-sensitive customer feedback system which are found by extending an established DSR methodology with value-sensitive design methods; 
II) an effective software artifact in terms of useability and user acceptance, which is evaluated in both a four-day field experiment with a large Swiss library and 132 of its customers and in a focus group with experts and managers from this library; 
III) a demonstration how blockchain technology can afford three design principles of effective customer feedback systems; IV) theoretical implications for DSR that are derived from a focus on value-sensitive design. Such a focus can  i) reduce the design space of the IT artefact and thus make the design more efficient in terms of required time and ii) explicitly provide the rationale for design principles in form of values.

This paper is organized as follows: In Section \ref{sec:lit_review}, a literature review about CFS in organizations is given. The DSR methodology that this paper follows is illustrated in Section \ref{sec:methodology}, while the findings and artifacts from applying this methodology are outlined in Section \ref{sec:findings}. Thereafter, Section \ref{sec:discussion} embeds the findings as design knowledge chunks \citep{vom2020special} into the broader research journey and illustrates the finalized design principles for customer feedback systems. 
Finally, in Section \ref{sec:conclusion} a conclusion is drawn and an outlook on future work is given.

\section{Research Background}
\label{sec:lit_review}

Customer feedback is an important element of an organizations' quality management \citep{chase_1991}, as the perceived quality of services and products is related to market share and return on investment \citep{Parasuraman_1985}. This is particularly relevant for service businesses \citep{chase_1991} such as libraries \citep{casey2006library}, since there is an increased emphasis on service quality rather than on manufacturing quality \citep{Vargo_2004}. This importance is recognized within the seminal work of  \citet{Sampson_1999}, who developed a framework for designing customer feedback systems (CFS) to improve service quality. 
Furthermore, an overview of advantages and disadvantages of different feedback collection systems and their designs for improving service quality in organizations have been illustrated by \citet{wirtz_2000}.

Usually, in these systems, customers provide feedback on services in the form of online reviews either directly on the selling platform (e.g. rating a service booked on Fiverr\footnote{Fiverr is an online marketplace for freelance services: https://www.fiverr.com/ (last accessed 2021-12-10)}) or on specific review platforms (e.g. providing travel reviews on Tripadvisor\footnote{Tripadvisor is an online travel company that operates a website and mobile app with user-generated content and comparison shopping website: https://www.tripadvisor.com/ (last accessed 2021-12-10}) \citep{Schneider_2020}. Although online ratings do not necessarily provide an objective measure of service quality \citep{Langhe_2015}, they are highly influential for customer-decision making, which is reflected in sales and consequently in business success \citep{Simonson_2016}. Customers' online rating data can even be utilized to predict service business failures months in advance as it has been shown for the hospitality industry \citep{Naumzik_2021}.
Despite their high relevance for influencing customers' decision making, online ratings are potentially challenged by various factors including self-selection \citep{Hu_2009}, social influence \citep{Muchnik_2013}, manipulation of reviews \citep{zhuang2018manufactured, gossling2018manager}, and dimensional rating \citep{Schneider_2020}. Particularly, single-dimensional rating scales (e.g. Google reviews, which allow for a score from 1 to 5 stars) are not suitable to assess complex performance dimensions \citep{ittner_2003}. By these means, an organization will only receive a single rating, which often cannot be associated with a particular service that the customer received. To overcome this issue, firms have to invest in dedicated CFS, which allow them to receive feedback that they can benefit from (e.g. feedback on a specific service that they intend to improve) \citep{Sampson_1999}. Within such a system, the feedback is processed in the following sequential manner: channeling (i.e. feedback reception), processing (i.e. using the feedback for improvements), and conversion of the feedback into organization-wide knowledge \citep{BIRCHJENSEN_2020}. In this context, channeling is highly relevant, since feedback in a certain quantity and quality needs to be received to enable the subsequent steps of processing and conversion \citep{Lafky_2020}. Depth and extremity of reviews have been identified as useful indicators of the quality of feedback on an e-commerce platform \citep{Mudambi_2010}, which are found to be rather incentivized by social norms than financial rewards \citep{Burtch_20218}. Feedback quantity is known to be positively influenced by financial incentives \citep{Burtch_20218}, along with several other factors such as trust \citep{celuch2011search} and perceived usefulness (i.e. customers thinks that their feedback is useful for the organization) \citep{robinson_2013}.
Nevertheless, feedback quality may be reduced by applying such incentives \citep{Lafky_2020}. For instance, financial incentives lead to a reduction in feedback quality measured in depth (e.g., the length of a written review) while increasing feedback quantity measured in breadth (number of provided reviews) \citep{Burtch_20218}, thus revealing a trade-off between quantity and quality that is steered by the chosen incentive \citep{Lafky_2020}. 

Multi-dimensional incentives in the form of blockchain-based tokens have been proposed as an alternative to such financial incentives improving the properties of incentivized behavior such as actions contributing to sustainability \citep{kleineberg2021social,dapp2019toward,ballandies2021, dapp2021finance}.
In this regard, blockchain-based incentives have been suggested to improve the data quality in inter-organizational information exchange \citep{zavolokina2018incentivizing, hunhevicz2020incentivizing}. For instance, it has been found that Blockchain technology could contribute to trustworthy CFS in the tourism industry \citep{onder2018blockchain}. \citet{Chandratre_2019, Rahman_2020, gipp2017cryptsubmit} are among the first to propose and implement blockchain-based feedback systems. Although these systems utilize blockchain technology for tracking and the immutable storing of feedback items, they do not explore incentivizing feedback provision with blockchain-based tokens. This is a missed opportunity as, on the on hand, cryptoeconomic incentives carry monetary value \citep{Kranz_2019, sunyaev2021token}, and thus could motivate users to increase feedback quantity, while, on the other hand, they have different characteristics to money \citep{kleineberg2021social,dapp2019toward,ballandies2021, dapp2021finance}, and thus might impact feedback quality differently when compared to monetary incentives.  
In general, cryptoeconomic incentives in the form of blockchain-based tokens have been utilized to improve information sharing scenarios \citep{ballandies2022to}. Nevertheless, most approaches only utilize a single token incentive and do not consider the combination of two, which is a missed opportunity because this might improve system performance \citep{ballandies2022to}.

In order to construct such blockchain-based systems, Design Science Research (DSR) methods \citep{hevner2004design,hevner2010design,vom2020special} have been successfully applied within the IS community \citep{ballandies2021,Ostern_2020}. 
For this, amongst others, the model of \citet{zargham2018} is utilized that describes a system in five layers, as illustrated in Figure \ref{fig:system_design}. Design principles can only be explicitly formulated and implemented in the software system \citep{ballandies2021}, i.e. the bottom three layers of the model ((I-III) in Figure \ref{fig:system_design}). The upper two layers emerge and cannot be explicitly defined by the system designer (e.g. the associated researchers).

In summary, the following observations can be made about current CFS applied in organizations:
First, initial, prior findings provide promising evidence regarding the utilization of (multiple) blockchain-based tokens for incentivizing behavior, such as influencing the provision of high-quality feedback. However, they have not been incorporated within a feedback system and studied within a real-world use case. Second, research on CFS mostly focuses on hospitality, tourism, and e-commerce applications, while neglecting other application domains such as a library ecosystem.  
Third, CFS often focus on uncontextualized and solicited feedback in the form of single-dimensional rating scales.
In this work, we address these gaps by investigating the design of a value-sensitive blockchain-based feedback system that incentivizes the provision of feedback with multiple tokens and utilizes the concept of contextualization of feedback to enable users to increase the depth of their feedback and consequently feedback quality \citep{Burtch_20218}.  For this, we apply an established DSR methodology \citep{sonnenberg2012evaluations,hevner2004design,peffers2007design} and combine it with value-sensitive design methods \citep{friedman2013value,van2015handbook} that consists of expert interviews, stakeholder and value analysis, focus groups, and a four-day ethics commission approved socioeconomic experiment, involving a major Swiss library and its customers.

\section{Research Design}
\label{sec:methodology}

\begin{figure}[tb]
    \centering
    \includegraphics[width=0.9\textwidth]{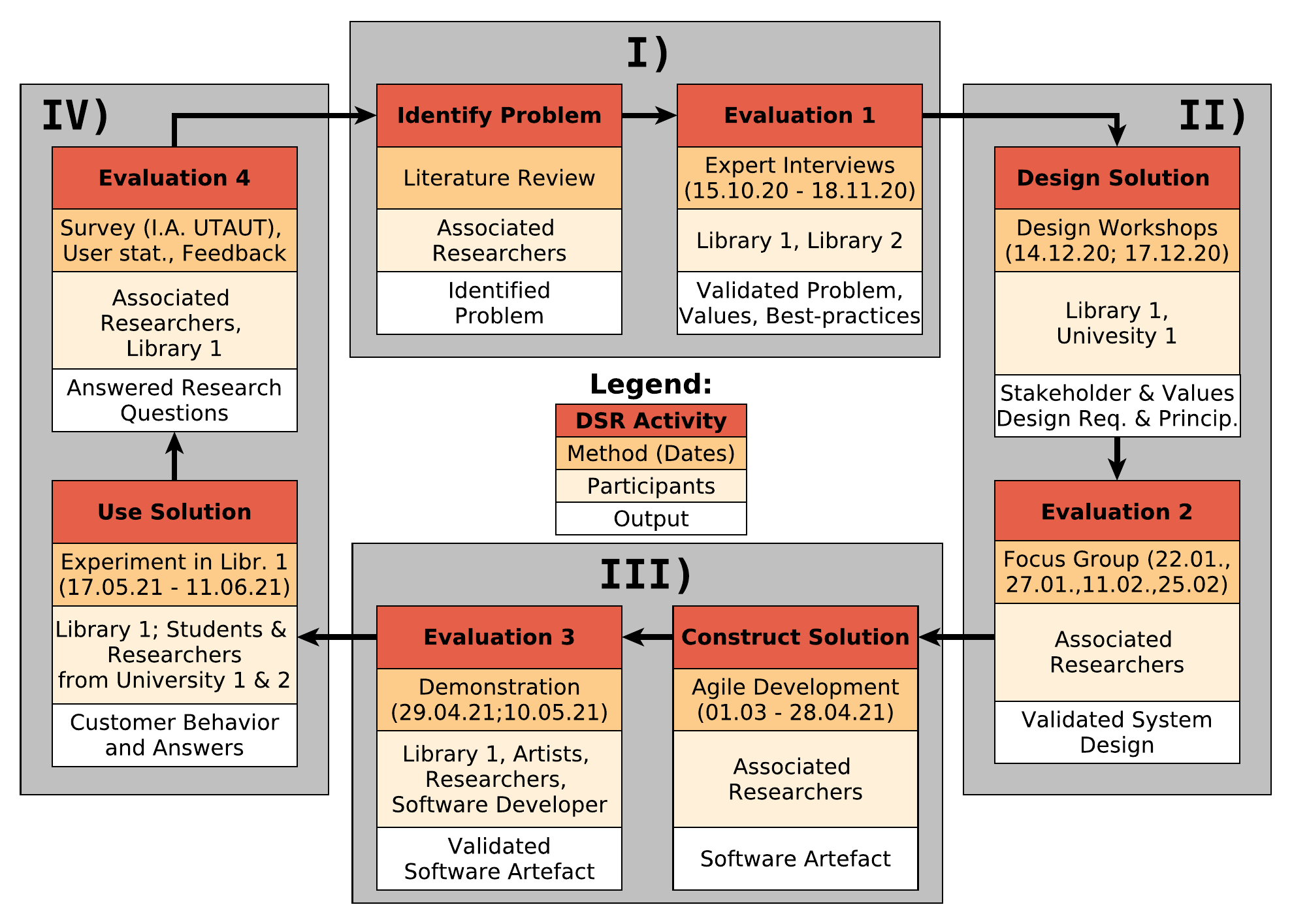}
    \caption{Activities, methods, participants and outputs of the four steps (I-IV) of the cyclic DSR process \citep{sonnenberg2012evaluations}}
    \label{fig:dsr_process}
\end{figure}

This research applies a DSR methodology \citep{hevner2010design,hevner2004design,peffers2007design} and combines it with value-sensitive design methods \citep{friedman2013value,van2015handbook}. We report on a DSR journey \citep{vom2020special}, that comprised of four iterations of concurrent design and evaluation \citep{sonnenberg2011evaluation,sonnenberg2012evaluations}. 
Figure \ref{fig:dsr_process} illustrates how the process with its four steps (I-IV) has been implemented in this work: 
In Step I, a literature review ("Identify Problem" in Figure \ref{fig:dsr_process}) identifies suboptimal feedback flows to and within organizations. Expert Interviews ("Evaluation 1" in Figure \ref{fig:dsr_process}) with employees of two major libraries evaluate this problem. Moreover, these interviews are utilized to identify important values and best-practice mechanisms in the context of feedback provision in these organizations that can be utilized to mitigate the identified problems and that are implemented within the constructed software artifact (Section \ref{sec:construct_solution}).  
In Step II, two design workshops with library employees and a customer of that library from University 1 resulted in (i) a stakeholder analysis which informs (ii) a value analysis which in turn facilitates the identification of (iii) design requirements ("Design solution" in Figure \ref{fig:dsr_process}). 
The requirements are incorporated in a system design via design principles and evaluated by the associated researchers ("Evaluation 2" in Figure \ref{fig:dsr_process}). In Step III, the associated researchers implemented the system by the means of agile development into a software artifact ("Construct solution" in Figure \ref{fig:dsr_process}). This artifact is validated in two demonstrations ("Evaluation 3" in Figure \ref{fig:dsr_process}) with focus groups consisting of library employees, researchers, software developers, and artists ("FG2" in Table \ref{tab:participants}). Finally in Step IV), the software artifact is put into use in an organizational context of Library 1 in the form of an experiment ("Use Solution" in Figure \ref{fig:dsr_process}) involving employees and customers of the library. The user behavior and answers to surveys are analyzed by the associated researchers ("Evaluation 4" in Figure \ref{fig:dsr_process}). Moreover, the collected feedback is evaluated by a focus group consisting of experts and executives of Library 1 ("FG3" in Table \ref{tab:participants}). 
\begin{figure}[tb]
    \centering
    \includegraphics[width=0.75\textwidth]{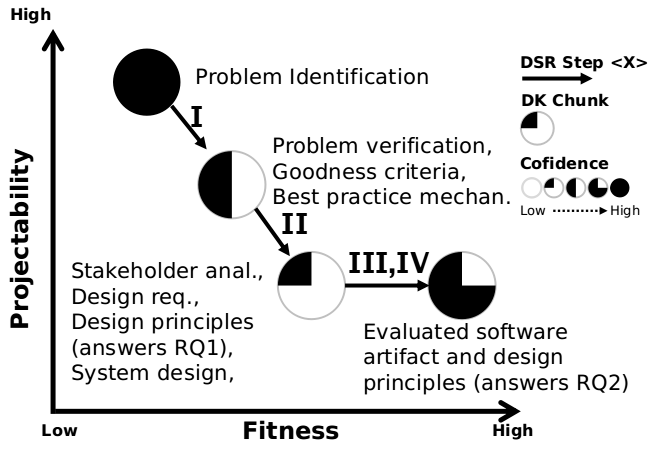}
    \caption{The contributions of the four steps of the research methodology (I-IV in Figure \ref{fig:dsr_process}) with regards to projectability (of the research context to new research contexts), fitness (of solving the target problem) and confidence (in the evaluation of the solution) of the design knowledge (DK) chunks of the research process as introduced in \citet{vom2020special}}
    \label{fig:dk_journey}
\end{figure}
Figure \ref{fig:dk_journey} illustrates how these four steps are positioned in the conducted DSR journey \citep{vom2020special} by connecting the design knowledge chunks obtained from each step of the DSR methodology (Figure \ref{fig:dsr_process}). 

\section{Research Findings}
\label{sec:findings}
In the following, we present the findings and artifacts obtained at each activity of the cyclic DSR process (Figure \ref{fig:dsr_process}).

\subsection{Identify Problem}
\label{sec:identify_problem}

During the first activity, expert interviews were conducted by the first author (A1) that evaluated the identified problem of suboptimal feedback flows in organizations (Section \ref{sec:introduction}). Table \ref{tab:participants} illustrates the twelve participants from two libraries who were interviewed in a semi-structured format. The participants were sampled by convenience based on the criterion that they work in a library. The interview protocol and guide are illustrated in Section 1 of the Supplementary Material (Table 1 and 2 of the Supplementary Material). Library 1 and 2 are among the largest in the German speaking countries. 
From Library 1 employees from all hierarchy levels (employee, team lead, section lead, director) and five of seven organizational sections were interviewed, including the director (ID 03, Table \ref{tab:participants}), whereas from Library 2 the director has been interviewed (ID 09, Table \ref{tab:participants}). 
The interviews were transcribed by a third party following the standard of \cite{Dresing2010} and were coded by the A1 with $19$ codes (Table 6 of the Supplementary Material). The codes were developed by the A1 and A2 following the method of \citet{o2020intercoder}. The definitions of the codes are given in Table 5 of the Supplementary Material. The interviews are then analyzed by grouping manually similar contents of each code into clusters (e.g., Figure 1 and Figure 2 of the Supplementary Material for the codes "status quo: challenge" and "risk").

\begin{table}[]
\caption{Participants in the Interviews (IW), Design Workshops (DW) and Focus Groups (FG1, FG2, FG3), their working area and hierarchy in their institution.}
\begin{tabular}{lllll} \hline
\textbf{ID} & \textbf{Hierarchy} & \textbf{Working Area}      & \textbf{Institution} & \textbf{Attended}             \\ \hline
01          & Team Lead          & Market Research                 & Library 1            & IW, DW, FG2, FG3 \\
02          & Employee           & Innovation \& Networking   & Library 1            & IW, DW, FG2, FG3 \\
03          & Director           & Upper Management & Library 1            & IW                    \\
04          & Employee           & Innovation \& Networking   & Library 1            & IW, DW, FG2, FG3 \\
05          & Team Lead          & Innovation \& Networking   & Library 1            & IW, DW, FG2, FG3 \\
06          & Section Lead       & IT-Services                & Library 1            & IW                     \\
07          & Section Lead       & Issuing Desk               & Library 1            & IW                     \\
08          & Employee           & Issuing Desk               & Library 1            & IW                     \\
09          & Director           & Upper Management & Library 2            & IW                     \\
10          & Team Lead          & Knowledge Management       & Library 1            & IW                     \\
11          & Employee           & Research data management   & Library 1            & IW, DW \\
12          & Team Lead          & E-Publishing               & Library 1            & IW                    \\
13          & Research Lead      & Information Systems                 & University 1         & DW        \\
14          & Artist              & Video Game Designer & - & FG1, FG2 \\
15          & Artist            & Interactive Media Designer & - & FG1, FG2 \\
16          & Researcher        & Neuroscience &University 3 & FG1, FG2 \\
17          & Software Developer & Machine Learning & Library 1 & FG1 \\
\hline     
\end{tabular}
\label{tab:participants}
\end{table}

In Section \ref{sec:problem}, a subset of these challenges are illustrated that are addressed in this work. Moreover, the interviews are utilized to identify the values stakeholders have in the context of feedback provision (Section \ref{sec:values}). Finally, the interviews are utilized to identify best practices when collecting feedback in the organizations (Section \ref{sec:best_practice_mechanisms}).

\subsubsection{Suboptimal Feedback Flows}

\label{sec:problem}
In order to obtain a multi-faceted perspective of the current and future challenges that exist or might arise with regard to feedback in an organizational context, the codes "status quo: challenge" and "risk" (Table 6 of the Supplementary Material) are analyzed. A comprehensive compilation of challenges are identified and illustrated in Figure 1 and Figure 2 of the Supplementary Material. The following challenges are addressed in this work: 
i) Mobilising non-customers: Obtaining feedback from those that do not utilize the library services, respectively those that are unaware that they are utilizing a service of the library is difficult. (ii) Hierarchy of the organization prevents agile processing of feedback: Feedback is not forwarded preventing the recognition of the feedback by the responsible organizational unit (iii) Difficulty to distinguish important from unimportant feedback: When the quantity of collected feedback is high and enters the organization via various channels and organizational units, identifying feedback that would result in an improvement of services is hard. (iv) Difficulty to evaluate the quality of questions utilized in solicited feedback: When designing surveys, often similar questions are repeated or the posed questions are not evaluated if they enable a comprehensive answer (e.g., a limited set of answer options in single-choice type questions).
(v) Monetary incentives may increase quantity while reducing quality of collected feedback when awarded to customers for the provision of feedback.

\subsubsection{Values}
\label{sec:values}

The top six mentioned values are Anonymity (17), Transparency (16), Openness (11), Safety (11), Real-world Human Interactions (10), and Simplicity (10) (Table 7 of the Supplementary Material lists all mentioned values). The mentioning of anonymity and transparency are positively biased by the interview guide (Table 3 and 4 of the Supplementary Material) by surveying participants about the importance of those values. Openness, Safety, Real-world Human interactions, and Simplicity have been brought to the discussion by the experts. In particular, the value of real-world human interactions is considered as important to facilitate a successful feedback process as it enables the exchange of informal feedback: It facilitates the recognition of gestures and the exchange of spontaneous and unsolicited feedback.

\subsubsection{Best Practice Mechanisms} 
\label{sec:best_practice_mechanisms}

Six best practice mechanisms that the library has established have been identified (Section 1.5.1 of the Supplementary Material). The two that are integrated in the software artifact of this workd: i) A web-based feedback wall in the form of a pin-board is utilized where users can quickly post around the clock unsolicited feedback. The wall enables anonymous input and open/ transparent visibility of feedback items. The newest feedback is always on top. ii) Physical feedback collection points in the form of boxes are installed in the library buildings which facilitated the collection of high quantity feedback.

\subsection{Design Solution}
\label{sec:design_solution}

During this activity, two design workshops are conducted with employees of Library 1 and a customer of the library from University 1 (DW in Table \ref{tab:participants}) to identify the design requirements of the feedback system (Figure \ref{fig:dsr_process}). A1 moderated the workshop, whereas A2 facilitated the technical setup in the background and assisted participants in case of questions. Due to the Covid-19 policies at the research institute, the workshops are conducted virtually utilizing zoom\footnote{A video conferencing platform: https://zoom.us/, last accessed: 2021-10-04} and Miro\footnote{Online whiteboard and visual collaboration platform: https://miro.com/, last accessed: 2021-10-04}. 

Table 8 of the Supplementary Material illustrates the workshop activities: Activities are either collaborative if participants interact with each other, or individual if participants have no interactions. At the beginning of workshop~1, all participants conducted a collaborative training to familiarize themselves with Miro. Because status differences are evident in the group and participants (partly) did not know each other beforehand which limits social interactions, both workshops utilized Brainwriting \citep{vangundy1984brain} as an individual Brainstorming method to identify the stakeholders and design requirements. 
Both, the rating of stakeholder interest and influence, and the association of these stakeholders with values are performed individually and were analyzed by the research team after the workshops:
By averaging participants' rating of stakeholders influence and interest in the solution, the final stakeholders influence / interest matrix and its clusters are identified (Figure 7 of the Supplementary Material). This summarized stakeholder map was accepted by all participants in the second workshop.
In order to facilitate that the stakeholders' values are accounted for in the design requirements of the system, the value analysis is performed as an intermediate step between the stakeholder analysis and the requirements elicitation: Each participant individually assigned all relevant values to a subset of stakeholders (Cluster 3 and 4 in Figure 7 of the Supplementary Material) of the overall stakeholder analysis. The strength of stakeholders' association with a value is illustrated in Table \ref{tab:values_per_stakeholder}. 
The values have been taken from value sensitive design literature \citep{harbers2017value,van2015conflicting,friedman2020value,huldtgren2015design,hanggli2021human}. 
The value of Excellence has been added by the participants during the first Design Workshop.

The final design requirements (Figure \ref{fig:design_requirements}) are then elicited by first identifying the requirements associated with each value via brainwriting and then to prioritize those requirements (Average $\geq 1$ in Table \ref{tab:values_per_stakeholder}). In order to familiarize the workshop participants with these values, a preliminary value association task has been performed with the participants at the end of workshop 1 to prepare them for the second workshop. 
The chosen brainwriting and clustering approach was tested before the workshops within a diverse focus group consisting of artists, a researcher, and a software developer (FG1 in Table \ref{tab:participants}). 

In the following, the outputs of the two Design Workshops are illustrated: a Stakeholder map (Section \ref{sec:stakeholder_analysis}), a ranking of values based on their average importance (Section \ref{sec:value_analysis}) and value-based design requirements (Section \ref{sec:design_requirements}). Moreover, elicited by the associated researchers from the design requirements, design principles that guide the construction of value-sensitive feedback systems (Section \ref{sec:design_principles}) and a system design based on these principles (Section \ref{sec:system_design}) are illustrated. 
\subsubsection{Stakeholder Analysis}
\label{sec:stakeholder_analysis}
Figure 7 of the Supplementary Materials illustrates the participants and interest groups of the feedback system in the form of a Stakeholder map \citep{rossner2018prospects}.

Four clusters are identified: Cluster (1) does not have a positive influence on the construction of the solution and a low to medium interest in it. The cluster contains stakeholders such as suppliers, other libraries, and publishers. Cluster (2) contains the legislation, politics, and public funding institutions. These stakeholders have an influence on the solution while not having a high interest in it. The interest in and possible influence on the solution of Cluster (3) is the highest. These stakeholders are key in the design requirements engineering of Section \ref{sec:design_requirements}. This cluster includes the management and experts of the library as well as average employees and customers (e.g. researchers and lecturers). The directorate of the library has the highest interest and influence in the solution. Cluster (4) contains potential users of the system (e.g. students) that have a high interest in the solution but a low influence on its design. As these stakeholders are potential users of the system, their perspective is considered in the design requirements engineering to positively influence the adoption of the solution. 

\subsubsection{Value Analysis}
\label{sec:value_analysis}

\begin{table}[]
\caption{Strength of stakeholder association for each value: Green (3) - strong, yellow (2) - medium, red~(1) - low, white (0) - none, sorted by average strength, as identified by the design workshop participants (Table \ref{tab:participants}). Above the dashed line are those values that received an average strength of $1$ and thus were considered in the requirements analysis. }
\begin{tabular}{lrrrrrrrrrrrrrrrcc}
\textbf{Value}       & \multicolumn{1}{l}{\rotatebox{90}{Project Lead}} & \multicolumn{1}{l}{\rotatebox{90}{Lecturers}} & \multicolumn{1}{l}{\rotatebox{90}{Researchers}} & \multicolumn{1}{l}{\rotatebox{90}{University Manag.}} & \multicolumn{1}{l}{\rotatebox{90}{Issue Desk Staff}} & \multicolumn{1}{l}{\rotatebox{90}{Specialist Team}} & \multicolumn{1}{l}{\rotatebox{90}{Marketing}} & \multicolumn{1}{l}{\rotatebox{90}{Marekt Research}} & \multicolumn{1}{l}{\rotatebox{90}{Mid. Management}} & \multicolumn{1}{l}{\rotatebox{90}{Product Manager}} & \multicolumn{1}{l}{\rotatebox{90}{Staff}} & \multicolumn{1}{l}{\rotatebox{90}{Directorate}} & \multicolumn{1}{l}{\rotatebox{90}{Legal Department}} & \multicolumn{1}{l}{\rotatebox{90}{Students}} & \multicolumn{1}{l}{\rotatebox{90}{Tech./ admin Staff}} & \multicolumn{1}{l}{\textbf{Avg.}}  & \multicolumn{1}{l}{\textbf{Var.}} \\
Credibility         & \cellcolor[HTML]{F4C7C3}1        & \cellcolor[HTML]{B7E1CD}3     & \cellcolor[HTML]{B7E1CD}3       & \cellcolor[HTML]{F4C7C3}1             & \cellcolor[HTML]{FCE8B2}2            & \cellcolor[HTML]{F4C7C3}1           & \cellcolor[HTML]{FCE8B2}2     & \cellcolor[HTML]{FCE8B2}2           & \cellcolor[HTML]{F4C7C3}1           & \cellcolor[HTML]{B7E1CD}3           & \cellcolor[HTML]{F4C7C3}1 & \cellcolor[HTML]{B7E1CD}3       & 0                                    & \cellcolor[HTML]{B7E1CD}3    & \cellcolor[HTML]{F4C7C3}1              & \cellcolor[HTML]{EFEFEF}\textbf{1.80} &\cellcolor[HTML]{EFEFEF}\textbf{1.03}   \\
Simplicity           & \cellcolor[HTML]{B7E1CD}3        & \cellcolor[HTML]{B7E1CD}3     & \cellcolor[HTML]{B7E1CD}3       & \cellcolor[HTML]{F4C7C3}1             & \cellcolor[HTML]{B7E1CD}3            & 0                                   & 0                             & \cellcolor[HTML]{F4C7C3}1           & \cellcolor[HTML]{F4C7C3}1           & \cellcolor[HTML]{F4C7C3}1           & \cellcolor[HTML]{F4C7C3}1 & \cellcolor[HTML]{FCE8B2}2       & 0                                    & \cellcolor[HTML]{B7E1CD}3    & \cellcolor[HTML]{B7E1CD}3              & \cellcolor[HTML]{EFEFEF}\textbf{1.67} & \cellcolor[HTML]{EFEFEF}\textbf{1.52} \\
Universal Useability & \cellcolor[HTML]{FCE8B2}2        & \cellcolor[HTML]{B7E1CD}3     & \cellcolor[HTML]{B7E1CD}3       & \cellcolor[HTML]{F4C7C3}1             & \cellcolor[HTML]{FCE8B2}2            & \cellcolor[HTML]{F4C7C3}1           & \cellcolor[HTML]{F4C7C3}1     & \cellcolor[HTML]{FCE8B2}2           & \cellcolor[HTML]{F4C7C3}1           & \cellcolor[HTML]{FCE8B2}2           & \cellcolor[HTML]{F4C7C3}1 & 0                               & 0                                    & \cellcolor[HTML]{B7E1CD}3    & \cellcolor[HTML]{B7E1CD}3              & \cellcolor[HTML]{EFEFEF}\textbf{1.67} & \cellcolor[HTML]{EFEFEF}\textbf{1.10}\\
Excellence           & \cellcolor[HTML]{B7E1CD}3        & \cellcolor[HTML]{FCE8B2}2     & \cellcolor[HTML]{B7E1CD}3       & \cellcolor[HTML]{B7E1CD}3             & 0                                    & \cellcolor[HTML]{F4C7C3}1           & \cellcolor[HTML]{F4C7C3}1     & 0                                   & \cellcolor[HTML]{B7E1CD}3           & 0                                   & \cellcolor[HTML]{F4C7C3}1 & \cellcolor[HTML]{B7E1CD}3       & \cellcolor[HTML]{F4C7C3}1            & 0                            & \cellcolor[HTML]{FCE8B2}2              & \cellcolor[HTML]{EFEFEF}\textbf{1.53}& \cellcolor[HTML]{EFEFEF}\textbf{1.55} \\
Efficiency           & \cellcolor[HTML]{F4C7C3}1        & \cellcolor[HTML]{B7E1CD}3     & \cellcolor[HTML]{B7E1CD}3       & 0                                     & \cellcolor[HTML]{B7E1CD}3            & 0                                   & 0                             & 0                                   & \cellcolor[HTML]{F4C7C3}1           & \cellcolor[HTML]{B7E1CD}3           & \cellcolor[HTML]{F4C7C3}1 & \cellcolor[HTML]{B7E1CD}3       & \cellcolor[HTML]{F4C7C3}1            & \cellcolor[HTML]{F4C7C3}1    & 0                                      & \cellcolor[HTML]{EFEFEF}\textbf{1.33} & \cellcolor[HTML]{EFEFEF}\textbf{1.67}\\
Transparency         & \cellcolor[HTML]{B7E1CD}3        & \cellcolor[HTML]{B7E1CD}3     & \cellcolor[HTML]{B7E1CD}3       & \cellcolor[HTML]{F4C7C3}1             & \cellcolor[HTML]{F4C7C3}1            & \cellcolor[HTML]{F4C7C3}1           & \cellcolor[HTML]{F4C7C3}1     & \cellcolor[HTML]{B7E1CD}3           & 0                                   & 0                                   & 0                         & 0                               & \cellcolor[HTML]{F4C7C3}1            & 0                            & 0                                      & \cellcolor[HTML]{EFEFEF}\textbf{1.13} & \cellcolor[HTML]{EFEFEF}\textbf{1.55}\\
Identity             & 0                                & \cellcolor[HTML]{F4C7C3}1     & \cellcolor[HTML]{F4C7C3}1       & \cellcolor[HTML]{FCE8B2}2             & \cellcolor[HTML]{F4C7C3}1            & 0                                   & \cellcolor[HTML]{B7E1CD}3     & 0                                   & 0                                   & \cellcolor[HTML]{B7E1CD}3           & \cellcolor[HTML]{B7E1CD}3 & \cellcolor[HTML]{FCE8B2}2       & 0                                    & 0                            & 0                                      & \cellcolor[HTML]{EFEFEF}\textbf{1.07}& \cellcolor[HTML]{EFEFEF}\textbf{1.50} \\
Autonomy             & \cellcolor[HTML]{FCE8B2}2        & \cellcolor[HTML]{B7E1CD}3     & \cellcolor[HTML]{B7E1CD}3       & 0                                     & \cellcolor[HTML]{FCE8B2}2            & \cellcolor[HTML]{FCE8B2}2           & 0                             & 0                                   & 0                                   & 0                                   & \cellcolor[HTML]{F4C7C3}1 & 0                               & 0                                    & \cellcolor[HTML]{FCE8B2}2    & \cellcolor[HTML]{F4C7C3}1              & \cellcolor[HTML]{EFEFEF}\textbf{1.07} & \cellcolor[HTML]{EFEFEF}\textbf{1.35}\\
Safety               & 0                                & \cellcolor[HTML]{FCE8B2}2     & \cellcolor[HTML]{FCE8B2}2       & \cellcolor[HTML]{FCE8B2}2             & 0                                    & 0                                   & 0                             & 0                                   & \cellcolor[HTML]{F4C7C3}1           & 0                                   & \cellcolor[HTML]{F4C7C3}1 & 0                               & \cellcolor[HTML]{B7E1CD}3            & \cellcolor[HTML]{FCE8B2}2    & \cellcolor[HTML]{FCE8B2}2              & \cellcolor[HTML]{EFEFEF}\textbf{1.00}  & \cellcolor[HTML]{EFEFEF}\textbf{1.14}   \\
Real-w. Hum. Inter.    & \cellcolor[HTML]{F4C7C3}1        & \cellcolor[HTML]{B7E1CD}3     & 0                               & 0                                     & \cellcolor[HTML]{FCE8B2}2            & \cellcolor[HTML]{FCE8B2}2           & \cellcolor[HTML]{B7E1CD}3     & 0                                   & 0                                   & 0                                   & \cellcolor[HTML]{B7E1CD}3 & 0                               & 0                                    & \cellcolor[HTML]{F4C7C3}1    & 0                                      & \cellcolor[HTML]{EFEFEF}\textbf{1.00}  & \cellcolor[HTML]{EFEFEF}\textbf{1.57}   \\  \reddashedline
Accountability       & \cellcolor[HTML]{F4C7C3}1        & \cellcolor[HTML]{F4C7C3}1     & \cellcolor[HTML]{F4C7C3}1       & \cellcolor[HTML]{FCE8B2}2             & 0                                    & 0                                   & \cellcolor[HTML]{F4C7C3}1     & 0                                   & \cellcolor[HTML]{F4C7C3}1           & \cellcolor[HTML]{FCE8B2}2           & 0                         & \cellcolor[HTML]{FCE8B2}2       & \cellcolor[HTML]{F4C7C3}1            & 0                            & \cellcolor[HTML]{F4C7C3}1              & \cellcolor[HTML]{EFEFEF}\textbf{0.87} & \cellcolor[HTML]{EFEFEF}\textbf{0.55}\\
Responsibility       & \cellcolor[HTML]{F4C7C3}1        & 0                             & 0                               & \cellcolor[HTML]{F4C7C3}1             & \cellcolor[HTML]{B7E1CD}3            & \cellcolor[HTML]{B7E1CD}3           & 0                             & 0                                   & 0                                   & \cellcolor[HTML]{F4C7C3}1           & \cellcolor[HTML]{B7E1CD}3 & 0                               & 0                                    & 0                            & \cellcolor[HTML]{F4C7C3}1              & \cellcolor[HTML]{EFEFEF}\textbf{0.87}& \cellcolor[HTML]{EFEFEF}\textbf{1.41} \\
Informed Consent     & 0                                & \cellcolor[HTML]{FCE8B2}2     & \cellcolor[HTML]{FCE8B2}2       & \cellcolor[HTML]{F4C7C3}1             & 0                                    & 0                                   & 0                             & \cellcolor[HTML]{F4C7C3}1           & \cellcolor[HTML]{F4C7C3}1           & 0                                   & 0                         & \cellcolor[HTML]{F4C7C3}1       & \cellcolor[HTML]{B7E1CD}3            & 0                            & \cellcolor[HTML]{F4C7C3}1              & \cellcolor[HTML]{EFEFEF}\textbf{0.80} & \cellcolor[HTML]{EFEFEF}\textbf{0.89}  \\
Privacy              & \cellcolor[HTML]{F4C7C3}1        & \cellcolor[HTML]{F4C7C3}1     & 0                               & 0                                     & 0                                    & \cellcolor[HTML]{F4C7C3}1           & 0                             & 0                                   & 0                                   & 0                                   & \cellcolor[HTML]{B7E1CD}3 & 0                               & \cellcolor[HTML]{B7E1CD}3            & 0                            & \cellcolor[HTML]{B7E1CD}3              & \cellcolor[HTML]{EFEFEF}\textbf{0.80} & \cellcolor[HTML]{EFEFEF}\textbf{1.46}  \\
Inclusiveness        & \cellcolor[HTML]{F4C7C3}1        & 0                             & \cellcolor[HTML]{F4C7C3}1       & \cellcolor[HTML]{F4C7C3}1             & 0                                    & \cellcolor[HTML]{F4C7C3}1           & \cellcolor[HTML]{F4C7C3}1     & 0                                   & \cellcolor[HTML]{B7E1CD}3           & 0                                   & \cellcolor[HTML]{F4C7C3}1 & \cellcolor[HTML]{F4C7C3}1       & 0                                    & \cellcolor[HTML]{F4C7C3}1    & 0                                      & \cellcolor[HTML]{EFEFEF}\textbf{0.73} & \cellcolor[HTML]{EFEFEF}\textbf{0.64}\\
Freedom               & \cellcolor[HTML]{F4C7C3}1        & \cellcolor[HTML]{F4C7C3}1     & \cellcolor[HTML]{FCE8B2}2       & 0                                     & 0                                    & \cellcolor[HTML]{B7E1CD}3           & 0                             & 0                                   & 0                                   & \cellcolor[HTML]{F4C7C3}1           & \cellcolor[HTML]{F4C7C3}1 & 0                               & 0                                    & \cellcolor[HTML]{F4C7C3}1    & 0                                      & \cellcolor[HTML]{EFEFEF}\textbf{0.67}& \cellcolor[HTML]{EFEFEF}\textbf{0.81} \\
Openness              & 0                                & 0                             & \cellcolor[HTML]{F4C7C3}1       & 0                                     & \cellcolor[HTML]{F4C7C3}1            & 0                                   & 0                             & \cellcolor[HTML]{FCE8B2}2           & \cellcolor[HTML]{F4C7C3}1           & 0                                   & \cellcolor[HTML]{FCE8B2}2 & 0                               & \cellcolor[HTML]{F4C7C3}1            & \cellcolor[HTML]{FCE8B2}2    & 0                                      & \cellcolor[HTML]{EFEFEF}\textbf{0.67} & \cellcolor[HTML]{EFEFEF}\textbf{0.67}\\
Courtesy             & 0                                & 0                             & 0                               & 0                                     & \cellcolor[HTML]{B7E1CD}3            & \cellcolor[HTML]{F4C7C3}1           & 0                             & 0                                   & 0                                   & 0                                   & \cellcolor[HTML]{B7E1CD}3 & 0                               & 0                                    & 0                            & \cellcolor[HTML]{FCE8B2}2              & \cellcolor[HTML]{EFEFEF}\textbf{0.60}  & \cellcolor[HTML]{EFEFEF}\textbf{1.26} \\
Human Welfare        & 0                                & 0                             & 0                               & \cellcolor[HTML]{B7E1CD}3             & 0                                    & 0                                   & 0                             & 0                                   & \cellcolor[HTML]{F4C7C3}1           & \cellcolor[HTML]{FCE8B2}2           & 0                         & \cellcolor[HTML]{B7E1CD}3       & 0                                    & 0                            & 0                                      & \cellcolor[HTML]{EFEFEF}\textbf{0.60}  & \cellcolor[HTML]{EFEFEF}\textbf{1.26} \\
Sustainability       & 0                                & 0                             & \cellcolor[HTML]{F4C7C3}1       & 0                                     & 0                                    & \cellcolor[HTML]{F4C7C3}1           & \cellcolor[HTML]{B7E1CD}3     & 0                                   & 0                                   & \cellcolor[HTML]{F4C7C3}1           & 0                         & \cellcolor[HTML]{B7E1CD}3       & 0                                    & 0                            & 0                                      & \cellcolor[HTML]{EFEFEF}\textbf{0.60}& \cellcolor[HTML]{EFEFEF}\textbf{1.11}   \\
Calmness             & 0                                & \cellcolor[HTML]{F4C7C3}1     & 0                               & 0                                     & \cellcolor[HTML]{FCE8B2}2            & 0                                   & 0                             & 0                                   & 0                                   & 0                                   & 0                         & 0                               & \cellcolor[HTML]{B7E1CD}3            & 0                            & \cellcolor[HTML]{F4C7C3}1              & \cellcolor[HTML]{EFEFEF}\textbf{0.47}& \cellcolor[HTML]{EFEFEF}\textbf{0.84} \\
Trust                & \cellcolor[HTML]{F4C7C3}1        & 0                             & 0                               & 0                                     & \cellcolor[HTML]{FCE8B2}2            & \cellcolor[HTML]{F4C7C3}1           & 0                             & 0                                   & 0                                   & 0                                   & \cellcolor[HTML]{FCE8B2}2 & \cellcolor[HTML]{F4C7C3}1       & 0                                    & 0                            & 0                                      & \cellcolor[HTML]{EFEFEF}\textbf{0.47}& \cellcolor[HTML]{EFEFEF}\textbf{0.55} \\
Resilience           & \cellcolor[HTML]{F4C7C3}1        & 0                             & \cellcolor[HTML]{F4C7C3}1       & 0                                     & \cellcolor[HTML]{F4C7C3}1            & \cellcolor[HTML]{F4C7C3}1           & 0                             & 0                                   & \cellcolor[HTML]{F4C7C3}1           & 0                                   & 0                         & \cellcolor[HTML]{F4C7C3}1       & 0                                    & 0                            & 0                                      & \cellcolor[HTML]{EFEFEF}\textbf{0.40}  & \cellcolor[HTML]{EFEFEF}\textbf{0.26} \\
Ownership            & \cellcolor[HTML]{FCE8B2}2        & 0                             & 0                               & 0                                     & 0                                    & 0                                   & 0                             & 0                                   & 0                                   & \cellcolor[HTML]{B7E1CD}3           & 0                         & 0                               & 0                                    & 0                            & 0                                      & \cellcolor[HTML]{EFEFEF}\textbf{0.33} & \cellcolor[HTML]{EFEFEF}\textbf{0.81}\\
Unbiased             & 0                                & 0                             & \cellcolor[HTML]{F4C7C3}1       & 0                                     & 0                                    & 0                                   & 0                             & 0                                   & 0                                   & \cellcolor[HTML]{F4C7C3}1           & \cellcolor[HTML]{F4C7C3}1 & 0                               & 0                                    & 0                            & 0                                      & \cellcolor[HTML]{EFEFEF}\textbf{0.20} & \cellcolor[HTML]{EFEFEF}\textbf{0.17}  
\end{tabular}
\label{tab:values_per_stakeholder}
\end{table}

Table \ref{tab:values_per_stakeholder} illustrates which values the various Stakeholders from Cluster (3) and (4) carry, sorted by strength of association (output of the second Design Workshop).
The following values have the strongest association with the stakeholders (average strength $\geq 1$): Credibility, Simplicity, Universal Usability, Excellence, Efficiency, Identity, Autonomy, Safety, and Physical human interaction. These values are considered in the following sections (Section \ref{sec:design_requirements} and \ref{sec:design_principles}), which were only extended with the values of Human Welfare and Sustainability that are of importance to the directorate of the library (the most important stakeholder according to the analysis of Section \ref{sec:stakeholder_analysis}). Please refer to related work for an introduction of these values \citep{harbers2017value,van2015conflicting,friedman2020value,huldtgren2015design,hanggli2021human}.

\subsubsection{Design Requirements}
\label{sec:design_requirements}
\begin{figure}[tb]
    \centering
    \includegraphics[width=0.9\textwidth]{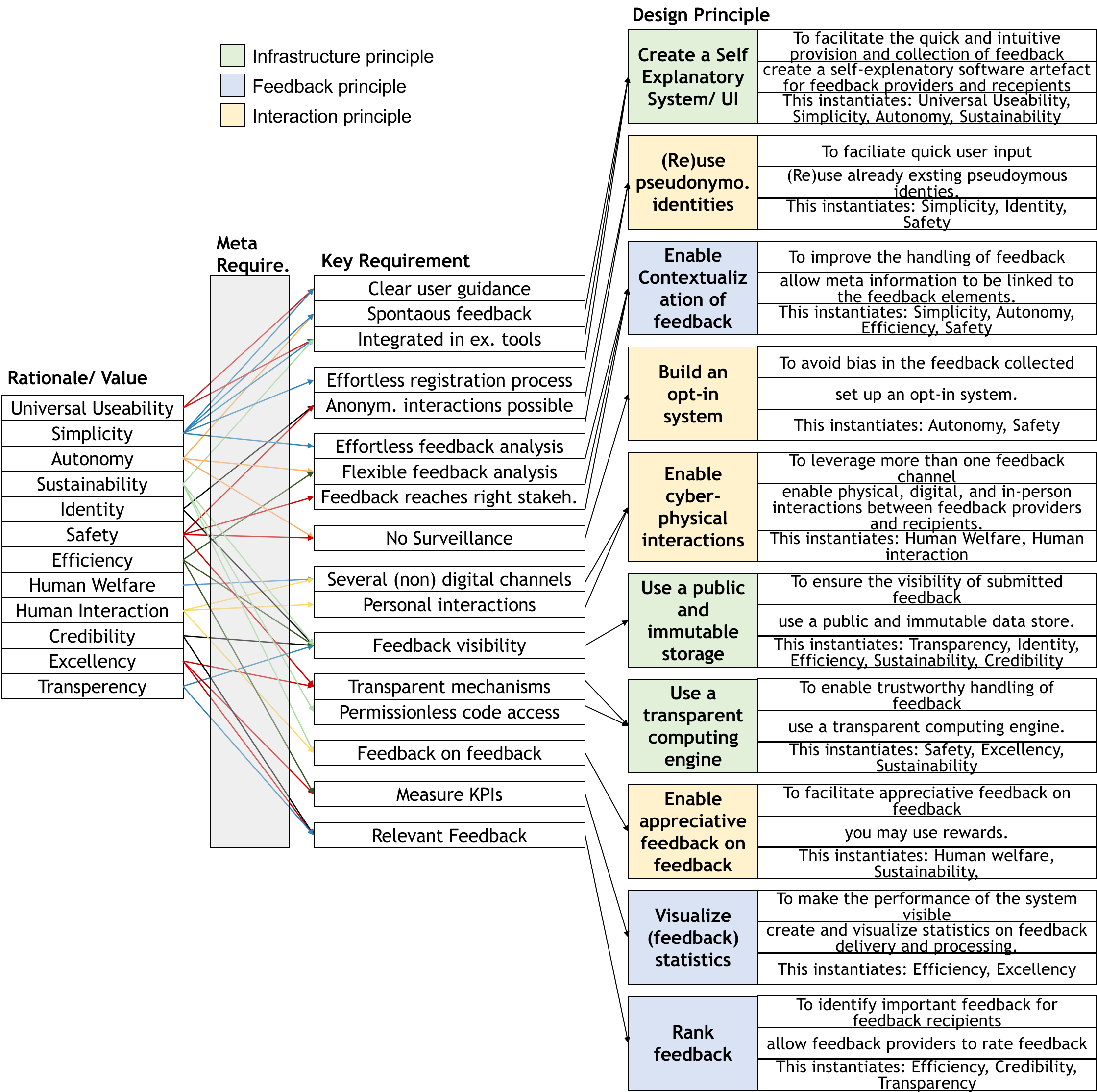}
    \caption{Identified values (left), requirements associated with these values (middle) and the design principles facilitating those requirements (right). The design principles utilize the framework of TBD for representation. The boundary condition for all principles are customer feedback systems. The rationale for a principle are the instantiated values (bottom box of each principle).}
    \label{fig:design_requirements}
\end{figure}

Figure \ref{fig:design_requirements} illustrates the identified important values (Section \ref{sec:value_analysis}) and the associated design requirements, which are the output of the second Design Workshop (Table \ref{tab:design_workshops}). Because the directorate is the stakeholder with the greatest influence on and interest in the solution (Section \ref{sec:stakeholder_analysis}), its values are added to the identified important values (Section \ref{sec:value_analysis}). In total 97 requirements are identified of which 32 are associated with the important values. The full list of requirements are given in Tables 12-14 of the Supplementary Materials.

\subsubsection{Design Principles}
\label{sec:design_principles}
By grouping the key requirements (Section \ref{sec:design_requirements}) into solution clusters, 10 design principles are identified (Figure \ref{fig:design_requirements}) that guide the construction of a value-sensitive feedback system. Three types of principles are found: i) infrastructure principles informing about technological requirements, ii) feedback principles illustrating the handling of the collected feedback, and iii) interaction principles illustrating the interplay among stakeholders and the system. Figure \ref{fig:design_requirements} depicts the 10 design principles in the framework introduced by \citet{gregor2020research} stating the aim, mechanism and rationale of each principle. The rationale are the values that inform the requirements which resulted in the principle. The context for each principle are customer feedback systems.
In the following, these principles are illustrated in greater detail.

\textbf{Infrastructure principles:} Feedback system designers should use a public and trustworthy storage infrastructure combined with a transparent computing engine to ensure the unconditional visibility of submitted feedback and its trustworthy post-processing. Moreover, software tools that are created should be self-explanatory to facilitate the quick and intuitive provision of collected feedback.

\textbf{Feedback principles:} Feedback items should be contextualized such that metadata\footnote{Metadata contributes to usability of information, a quality dimension of data \citep{cai2015challenges} which can be implemented utilizing semantic web technologies \citep{ballandies2021mobile}} of feedback (e.g., location of provision, the receiver, or importance) is also stored because this can improve the post-processing of feedback.
A possibility for ranking feedback to visualize the impact of each feedback item should be integrated to identify important feedback for feedback recipients. Also, the feedback should be aggregated and visualized in statistics to the stakeholders to make the performance of the system visible.

\textbf{Interaction principles:}
Stakeholders of the system should be enabled to have personal contacts in both, the cyberspace, but also in the physical reality to leverage on more than one feedback channel. Already existing pseudonymous identities could be (re)used to facilitate quick user input. Also, participation in the system should be voluntary in order to avoid bias in the feedback collected. Moreover, rewards could be utilized to facilitate appreciative feedback on feedback.

\subsubsection{System Design}
\label{sec:system_design}
\begin{figure}[tb]
    \centering
    \includegraphics[width=0.9\textwidth]{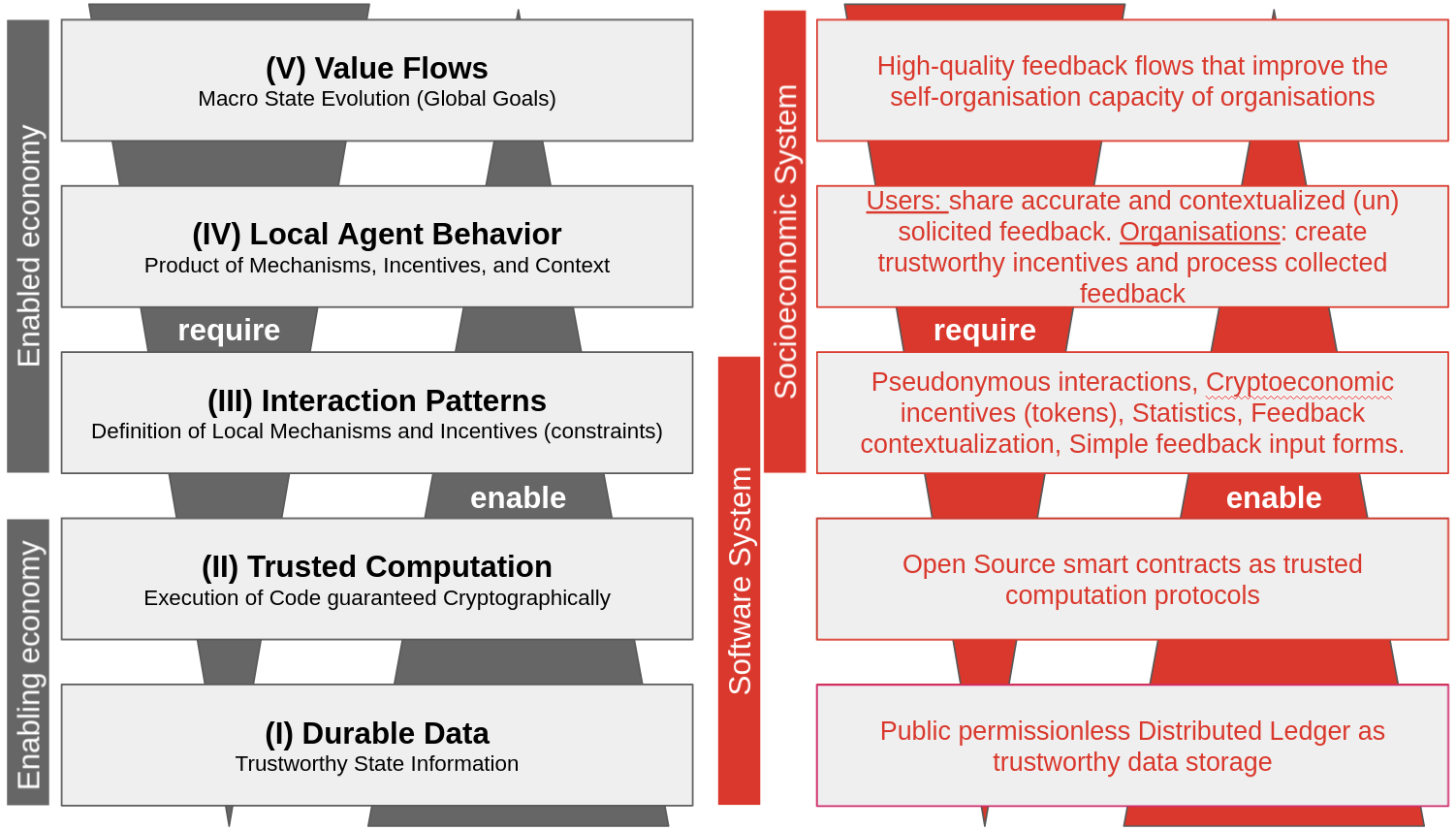}
    \caption{Utilizing the model of \citet{zargham2018} (left, grey), the interaction of the socio-economic feed4org system with its underlying software system is illustrated (red, right)  (adapted from \citet{ballandies2021}).}
    \label{fig:system_design}
\end{figure}

Figure \ref{fig:system_design} illustrates the system design that is found by instantiating the identified design principles.
The mapping of design principles to system design choices is illustrated in the following. Design principles that are related to the instantiation of the software artifact are illustrated in Section \ref{sec:construct_solution}.

\textbf{Distributed Ledger: }
A distributed ledger is a distributed data structure, whose entries are written by participants of a consensus mechanism after reaching an agreement on the validity of the entries. Such a consensus mechanism is called permissionless, if the public can participate in it, otherwise it is called permissioned \citep{ballandies2021decrypting}.
A permissionless distributed ledger (e.g. blockchain) affords a public and trustworthy storage (Layer I in Figure \ref{fig:system_design}, Design Principle 1 in Table \ref{tab:design_principles}), which in turn facilitates a transparent computing engine if open source smart contracts are utilized as trusted computation protocols (Layer II in Figure \ref{fig:system_design}, Design Principle 2 in Table \ref{tab:design_principles}). 
With these protocols, interactions patterns (Layer III in Figure \ref{fig:system_design}) can be implemented in the platform that define what users of the system can or cannot do \citep{ballandies2021}: (i) blockchain-based cryptoeconomic incentives in the form of tokens can be defined and awarded as rewards to users for the provision of high-quality feedback or respectful behavior (Design Principle 11 in Talbe \ref{tab:design_principles}) ; (ii) Also, diverse and trustworthy statistics can be aggregated and shown to the stakeholders of the system by analyzing the publicly accessible data storage via trusted computing protocols (Design Principle 6 in Table \ref{tab:design_principles}); (iii) Moreover, public distributed ledgers (e.g. Bitcoin) do not restrict the creation of user identities \citep{ballandies2021decrypting} (e.g. no know your customer policies), thus each user can decide how much information is revealed about their identity which facilitates pseudonymous interactions among stakeholders (Design Principle 9 in Table \ref{tab:design_principles}).

\begin{figure}
\begin{floatrow}
\ffigbox{%
  \includegraphics[width=0.5\textwidth]{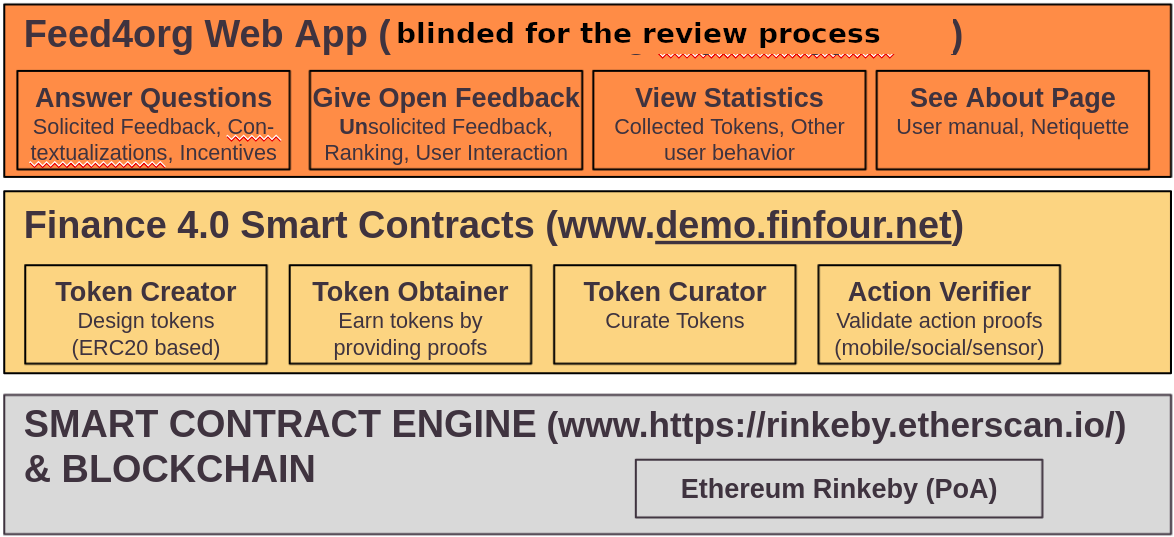}
}{%
   \caption{Software Stack of the feed4org app. }\label{fig:feed4org_architecture}
}
\ffigbox{%
  \includegraphics[width=0.5\textwidth]{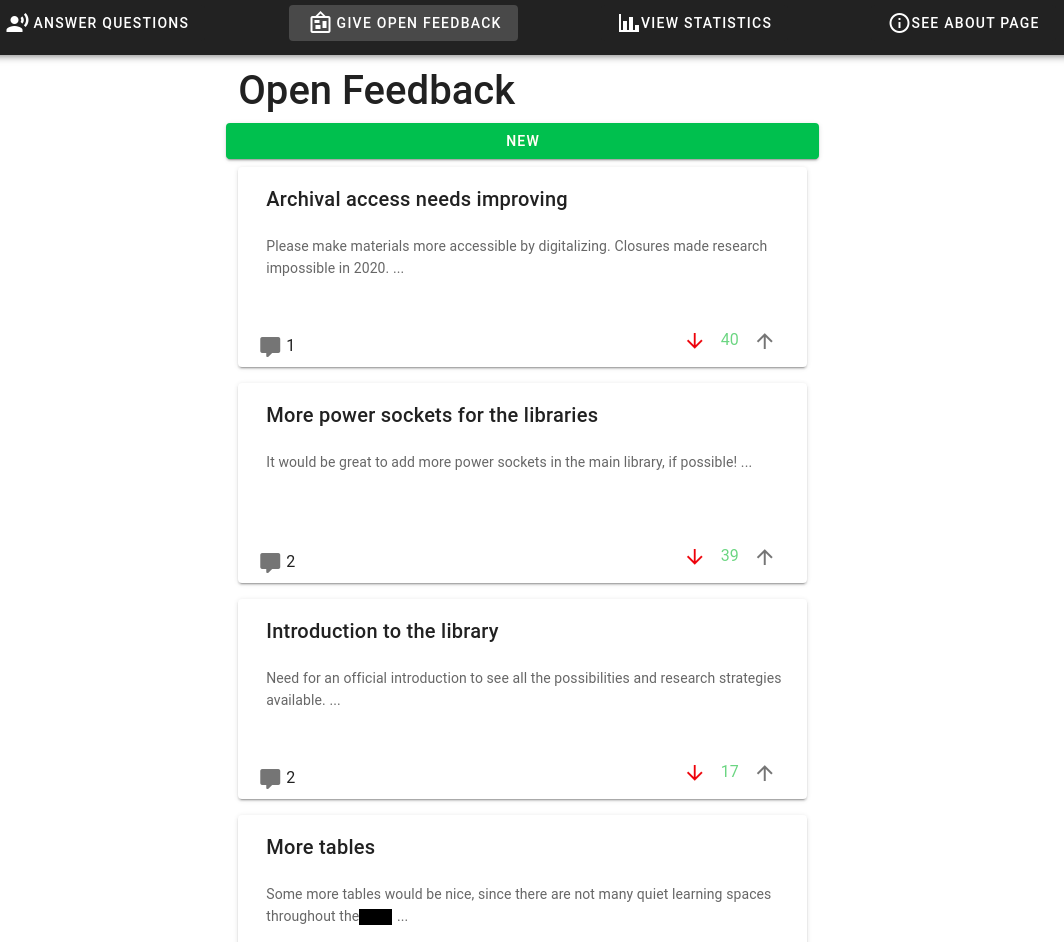}
}{%
   \caption{Open Feedback View (feedback wall) of the software artifact.}\label{fig:open_feedback_view}
}
\end{floatrow}
\end{figure}

\textbf{Contextualization and Incentives:}
In order to improve the processing and depth of the collected feedback, which is associated with feedback quality \citep{Mudambi_2010}, users are enabled to contextualize their feedback with meta-properties (Design Principle 4 in Table \ref{tab:design_principles}) such as the importance of the feedback, their satisfaction with the answer options, general comments, and the target audience to which their feedback is directed. Also, the system provides solicited (survey style questions, e.g. Figure \ref{fig:answer_fig}) and unsolicited (reddit style forum, e.g. Figure \ref{fig:open_feedback_view}) input forms such that diverse feedback can be collected  (Design Principle 7 in Table \ref{tab:design_principles}). 
The system encourages both, the provision of solicited feedback and the contextualization of this feedback, by incentivizing stakeholders with cryptoeconomic rewards in the form of tokens (Design Principle 11 in Table \ref{tab:design_principles}). 

\textbf{Cyber-physical interactions:}
Personal contact is enabled by following a cyber-physical approach: Stakeholders can interact via the created software artifact but are also encouraged to meet in the real-world\footnote{The real-world approach has been removed from the user study (Section \ref{sec:use_solution}) due to the COVID-19 restrictions at the time and location of the experiment.} at the library facilities by i) including interactive answer options such as taking photos at the library facility and ii) turning spots in the library into "real-time digital voting centers" \citep{hanggli2021human}, e.g. via the proof of witnessed presence \citep{pournaras2020proof}.

\subsection{Construct Solution}
\label{sec:construct_solution}

Utilizing agile development, a software artifact (referred to as \textit{feed4org app}) is constructed that is evaluated by demonstration in a focus group (Figure \ref{fig:dsr_process}): Based on the identified requirements (Section \ref{sec:design_requirements}), design principles (Section \ref{sec:design_principles}) and system design (Section \ref{sec:design_solution}), a software artifact is created to incentivize the provision of high-quality feedback to organizations. Figure \ref{fig:feed4org_architecture} illustrates the software stack: A web app is built using the VUEjs\footnote{https://vuejs.org/, last accessed: 2021-10-17} framework, on top of the Finance 4.0 software \citep{ballandies2021,ballandies2021finance}, which enables the creation of cryptoeconomic incentives in the form of tokens.
By utilizing the Finance 4.0 software stack, the trustworthy data storage and computation engine of the Ethereum blockchain is utilized that facilitates durable data and trusted computation as required by the system design (Section \ref{sec:system_design}, Design Principles 1 and 2 in Table \ref{tab:design_principles}). In particular, public blockchains, when compared to private blockhains, are i) transparent and publicly verifiable \citep{yang2020public}, and ii) secure \citep{yang2020public,ballandies2021decrypting}, thus incorporating values such as credibility and and safety, which are especially important for a public library (Figure \ref{fig:design_requirements}) offering services as a public good to engage customers, as identified in Section \ref{sec:value_analysis}. Also, utilizing a public blockchain reduces cost for the library organization as an own infrastructure does not need to be maintained \citep{yang2020public}. 
Moreover, Finance 4.0 facilitates the creation of tokens and proof verifications for awarding these tokens to feedback providers (Design Principle 11 in Table \ref{tab:design_principles}). By tailoring these incentives to library customers\footnote{please refer to Section \ref{sec:answer_questions} for details on the utilized incentives} to make only those participate in the feedback provision, high scalability facilitating a large number of feedback items per second from a potentially unrestricted user base is not required. For such a scalability requirement, utilizing either a permissioned/ private blockchain \citep{ballandies2021decrypting} or newer public blockchains (e.g. Solana \citep{yakovenko2018solana}) might be more suitable. 
The web app consists of four main components: i) Answer Question view that facilitates the provision of solicited feedback and the awarding of cryptoeconomic incentives (Design Principles 7 and 11 in Table \ref{tab:design_principles}), ii) Give Open Feedback view where users can provide unsolicited feedback (Design Principle 7 in Table \ref{tab:design_principles}), iii) View Statistics that informs users about their collected cryptoeconomic tokens and the behavior of others users (Design Principle 6 in Table \ref{tab:design_principles}) and iv) See About Page where information about the app and a Netiquette are displayed.
In the following, these components are illustrated in greater detail. 

\subsubsection{Answer Questions}
\label{sec:answer_questions}

\begin{figure}[tb]
    \centering
    \includegraphics[width=1.0\textwidth]{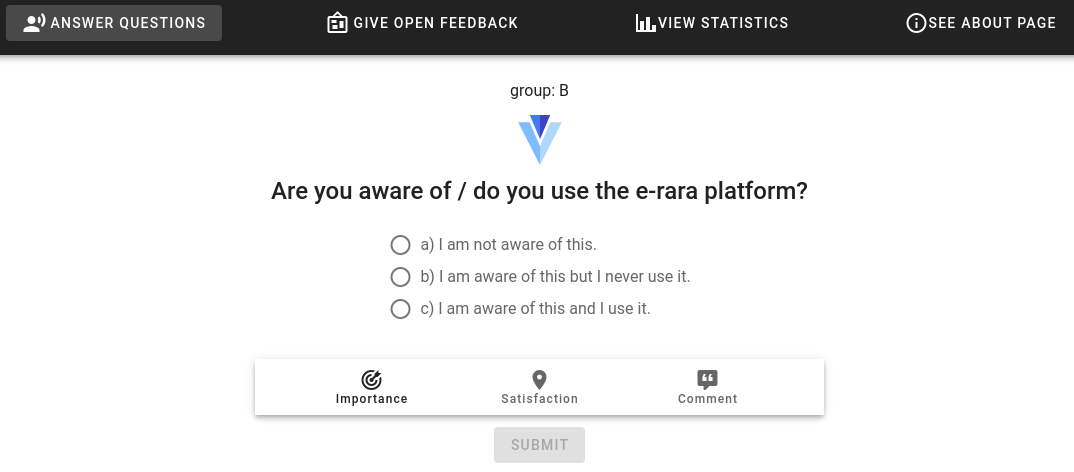}
    \caption{Answer View - Entry point to the feed4org app where the organization can ask for solicited feedback. Users have the possiblity to contextualize the feedback with its importance or their satisfaction. Moreover, each feedback item can be commented.}
    \label{fig:answer_fig}
\end{figure}

When users enter the app, they have the possibility to give feedback on questions posed by the library (Figure \ref{fig:answer_fig}). 
The following question types are implemented: Single-choice, multiple-choice, likert scales, open text, and combinations of these options. 

\textbf{Contextualization: }
In addition to answering questions, users can contextualize their answers (Design Principle 4 in Table \ref{tab:design_principles}) by clicking on the contextualization buttons (bottom buttons in Figure \ref{fig:answer_fig}). Three types of contextualizations are implemented. Users can state with the Importance contextualization how important the question is for the library to improve their services, with the Satisfaction contextualization how satisfied they are with the range of answer options and with the Comment contextualization to provide further comments in an open feedback form (Figure 10 of the Supplementary Material).

\begin{figure}
\begin{floatrow}
\ffigbox{%
  \includegraphics[width=\linewidth]{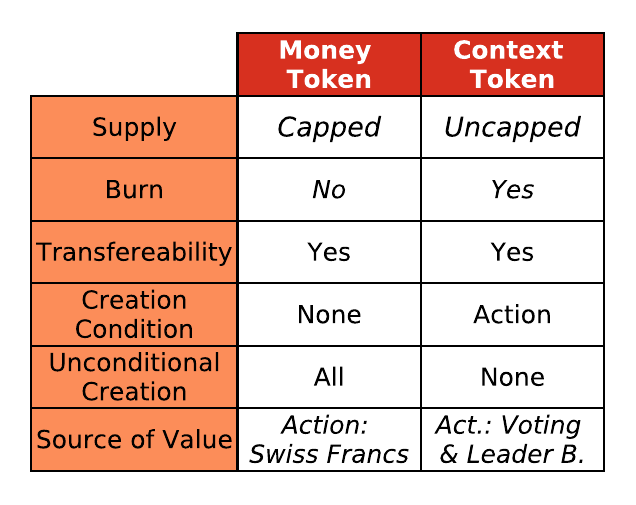}
}{%
  \caption{Token Design of the utilized cryptoeconomic incentives utilizing the DLT system taxonomy \citep{ballandies2021decrypting}.}\label{fig:token_design}
}
\capbtabbox{%

\begin{tabular}{cllcccc} \hline
\multicolumn{1}{l}{}                  &                      & \multicolumn{1}{c}{\multirow{2}{*}{\textbf{Total}}} & \multicolumn{2}{c}{\textbf{user}}                                                                                         & \multicolumn{2}{c}{\textbf{unaw.-user}}                                                                            \\
\multicolumn{1}{l}{}                  &                       & \multicolumn{1}{c}{}                                & \textit{\begin{tabular}[c]{@{}c@{}}\textit{C}\\ 18\end{tabular}} & \textit{\begin{tabular}[c]{@{}c@{}}\textit{T}\\ 54\end{tabular}} & \textit{\begin{tabular}[c]{@{}c@{}}\textit{C}\\ 15\end{tabular}} & \begin{tabular}[c]{@{}c@{}}\textit{T}\\ 45\end{tabular} \\ \hline
\parbox[t]{2mm}{\multirow{2}{*}{\rotatebox[origin=c]{90}{\multirow{2}{*}{\textbf{\begin{tabular}[c]{@{}l@{}}Feed-\\ back\end{tabular}}}}}}  & \textit{Solicited}    & \textit{21286}                                      & 71.5                                                        & 189.7                                                       & 63.9                                                        & 195.5                                              \\
                                      & \textit{Unsolicited}  & \textit{55}                                         & 0.8                                                         & 0.5                                                         & 0.2                                                         & 0.2                                                \\
                                    \hline
\parbox[t]{2mm}{\multirow{3}{*}{\rotatebox[origin=c]{90}{\multirow{2}{*}{\textbf{\begin{tabular}[c]{@{}l@{}}Con-\\ text \end{tabular}}}}}}    & \textit{Importance}   & \textit{6018}                                       & 28.4                                                        & 52.1                                                        & 33.5                                                        & 48.6                                               \\
                                      & \textit{Satisfaction} & \textit{5692}                                       & 26.9                                                        & 50.3                                                        & 22.5                                                        & 47.9                                               \\
                                      & \textit{Comments}     & \textit{2107}                                       & 15.2                                                        & 19.1                                                        & 9.3                                                         & 14.7                                               \\ \hline
\parbox[t]{2mm}{\multirow{4}{*}{\rotatebox[origin=c]{90}{\multirow{2}{*}{\textbf{\begin{tabular}[c]{@{}l@{}}Navigate\\ to\end{tabular}}}}}} & \textit{Question}     & \textit{6990}                                       & 19.9                                                        & 65.6                                                        & 15.8                                                        & 63.3                                               \\
                                      & \textit{Statistics}   & \textit{3605}                                       & 11                                                          & 35.9                                                        & 10.3                                                        & 29.2                                               \\
                                      & \textit{Open feed.}   & \textit{3094}                                       & 14.6                                                        & 30.3                                                        & 7.8                                                         & 23.9                                               \\
                                      & \textit{About View}   & \textit{549}                                        & 3                                                           & 5                                                           & 1.4                                                         & 4.5     \\ \hline                                           
\end{tabular}
}{%
  \caption{Total and mean amount of interactions with the software artifact for the 132 experiment participants and the treatment (T, with token incentives) and control group (C, no token incentives) for users and unaware-users.} \label{tab:app_interactions}
}
\end{floatrow}
\end{figure}

\textbf{Cryptoeconomic Incentivization} in the form of blockchain-based tokens is utilized via the Finance 4.0 platform to incentivize the provision of feedback. Figure \ref{fig:token_design} illustrates the design choices of the utilized cryptoeconomic incentives using a taxonomy for DLT systems \citep{ballandies2021decrypting}, as performed by \citet{dobler2019extension}:

The Money token is pre-mined and is pegged to the Swiss franc, thus being a stable coin collateralized with a fiat currency \citep{mita2019stablecoin}. Each token unit is worth 0.20 CHF. For each answered question, the user is rewarded with a token unit. 

The Context token is created whenever a contextualization action is performed and awarded to the feedback provider. Thus, this token is not capped, but its amount illustrates the number of contextualization actions performed in the system. Users can utilize this token to vote on the importance of unsolicited feedback (Section \ref{sec:give_open_feedback}, Design Principle 5 in Table \ref{tab:design_principles}). Moreover, the token is utilized to rank users in a leader board, which is displayed to all users in the View Statistics view of the app (Section \ref{sec:view_statistics}).

\subsubsection{Give Open Feedback}
\label{sec:give_open_feedback}

Via the Navigation bar (top bar in Figure \ref{fig:answer_fig}), users can switch to the Give Open feedback view. This view is based on the best practice mechanisms "feedback wall" of the library identified in Section \ref{sec:best_practice_mechanisms} and thus could be the integration point of this software artifact into the existing library software stack providing a low-threshold to provide feedback for existing library users familiar with that view (Design Principle 3 in Table \ref{tab:design_principles}). Via the feedback wall, users can provide unsolicited feedback to the organization (Design Principle 7 in Table \ref{tab:design_principles}). This paper extends this mechanism by i) enabling users to up and downvote a feedback item, facilitating the design principle of ranking feedback items (Design Principle 5 in Table \ref{tab:design_principles}), ii) to comment on a feedback item, facilitating personal contacts among users (Design principle 8 in Table \ref{tab:design_principles}), and iii) to provide area tags on a feedback item that connects it with strategic action areas of the library where the management of the library aims to improve their services (Design Principle 4 in Table \ref{tab:design_principles}).  
In order to up and downvote a feedback item, users are required to spend a unit of the context token (Section \ref{sec:answer_questions}).

\subsubsection{View Statistics}
\label{sec:view_statistics}
In the statistics view (Figure 11 of the Supplementary Material), users are informed about the amount of collected cryptoeconomic tokens. Moreover, the behavior of other users is displayed in a leaderboard that illustrates how many context tokens other users in the system collected. This facilitates the design principles of collecting statistics (Design Principle 6 in Table \ref{tab:design_principles}).

\subsection{Use Solution}
\label{sec:use_solution}

The software artifact is utilized in a four-day long real-time experiment to collect feedback from library customers. The collected feedback is then evaluated by both, statistical analysis, and a focus group consisting of library employees (FG3 in Table \ref{tab:participants}). 

The experiment has been conducted in collaboration with the Laboratory\footnote{
The reference reveals information identifying authors' affiliation. It will be added after the review process is finished.
}, who recruited the participants, guaranteed a fair compensation (10 CHF show-up fee, 30 CHF/h mean compensation), and facilitated anonymity for the participants by separating their identity information from the experiment data: The research team has only access to the latter.
For the four-day experiment setup\footnote{  
An Ethics commission approval has been obtained for the experiment (Section 2 of the Supplementary Material)}
 132 users were recruited in four waves receiving on avarage a compensation of 40.23 CHF. 
The software artifact and the experiment setup were evaluated with demonstrations before the experiment in two focus group meetings with library employees, artists, and a researcher (FG2 in Table \ref{tab:participants}).
The finalized experiment setup is composed as follows: At the beginning of each wave, participants obtained onboarding materials (Section 6.3.1 of the Supplementary Material) in which they are introduced to the app and answered demographic questions, Computer Self-Efficacy (adapted from \citet{compeau1995computer,thatcher2008internal} as performed in \citet{sun2019revisiting}) and Personal Innovativeness in IT (adapted from \citet{agarwal2000time} as performed in \citet{sun2019revisiting}) questions.

During the experiment phase, participants utilized the software artifact and provided feedback to Library 1. In the exit phase, users answered a questionnaire that included a UTAUT (Section 3 of the Supplementary Material, adapted from \citet{venkatesh2012consumer}) and questions regarding the value of cryptoeconomic tokens \citep{ballandies2021finance}. After the experiment, a focus group consisting of Library 1 employees (FG3 in Table \ref{tab:participants}) evaluated the quality of collected feedback.

In total, 132 participants completed the study with an average age of 23.2 years. Of these, 51.5 \% are male, 47.0 \% are female and 1.5 \% do not want to reveal their gender. On average, the participants self-report to be modest computer self-effective ($2.8$\footnote{\label{fn:self_efficacy} Average value on a 5-point likert scale (0 - strongly disagree to 4 - strongly agree).} computer self-efficacy \citep{sun2019revisiting}) and innovative in IT ($2.4$ innovativeness in IT \citep{sun2019revisiting}). Moreover, 54.5 \% of the participants are utilizing the services of the library that is focus of this study. 99 participants received the treatment in form of cryptoeconomic tokens.
In the following, user interactions with the App and user responses to the exit survey, and the evaluation of the focus group are illustrated.

\begin{table}[]
\caption{Mean, lower/ upper 95 \% confidence interval, standard deviation and number of participant answers of the constructs of effort expectancy (UTAUT, Table 15 of the Supplementary Material) and token value (Table 16 of the Supplementary Material) of Evaluation 4 that utilize a 5-point likert scale (0 - strongly diasagree to 4 - strongly agree). Token value is evaluated only for the treatment group as the control group did not utilize tokens.}
\begin{tabular}{lllllll} 
\textbf{Construct} & \textbf{QID}     & \textbf{Mean} & \textbf{CI (95\%) low} & \textbf{CI (95\%) up} & \textbf{Stdev} & \textbf{N} \\ \hline    \\
                   & UTAUT 5         & 3.00          & 2.87                   & 3.13                 & 0.78           & 132            \\
Effort             & UTAUT 6         & 2.77          &        2.62                & 2.91                  & 0.85           & 132            \\
Expectancy         & UTAUT 7          & 2.92          & 2.77                   & 3.07                  & 0.86           & 130            \\
                   & UTAUT 8          & 2.63          & 2.47                   & 2.79                  & 0.91           & 132            \\
                   & \textit{Average} & \textit{2.83} & \textit{2.68}          & \textit{2.98}         & \textit{0.85}  & \textit{131.5} \\ \hline \\
                   & FIN4 1          & 2.54          & 2.27                   & 2.81                  & 1.36           & 99           \\
Token              & FIN4 2           & 3.19          & 2.97                   & 3.42                  & 1.13           & 99            \\ Value
                   & FIN4 3           & 2.71          & 2.45                   & 2.97                  & 1.32           & 99        \\
                   & \textit{Average} & \textit{2.81} & \textit{2.56}          & \textit{3.07}         & \textit{1.27}  & \textit{99}   \\ \hline \\
\end{tabular}
\label{tab:utaut}
\end{table}

\subsubsection{App Interactions}
\label{sec:app_interaction}

Table \ref{tab:app_interactions} illustrates the user interactions with the software artifact. In total, 21286 solicited feedback items and 13817 contextualizations are collected from users which indicates a high useability of the artifact. In particular, the scoring on the UTAUT (Table \ref{tab:utaut}) for the effort expectancy (2.83) validates the design principle of simple user input/ self-explanatory UI (Figure \ref{fig:design_requirements}): Above all, it is easy for participants to learn using the software artifact (3.00). The focus group (FG4 in Table \ref{tab:participants}) indicated that this might be due to the clear focus in the design of the question answering and contextualization views which does not utilize unnecessary design elements. 

Table \ref{tab:usefulness} illustrates the usefulness evaluation of the artifact by the participants. Both, the treatment (2.82) and the control (2.80) group evaluate the features of the artifact as useful on average. In particular, the statistics view has been evaluated as most useful in the treatment group, whereas the control group rated the open feedback view as most useful. 

Figure 12 of the Supplementary Material illustrates the user interactions with the software artifact in a heatmap. The importance (6018) and satisfaction (5692) contextualization are more often utilized than the comment (2107) contextualization.

\begin{table}[]
\caption{Mean and variance of responses by the treatment (N=99) and control (N=33) group to the question "How useful did you find the following features of the app", utilizing a 5-point likert scale (0 - strongly diasagree to 4 - strongly agree).}
\begin{tabular}{lcccc} \hline 
                                        & \multicolumn{2}{c}{\textbf{Mean}} & \multicolumn{2}{c}{\textbf{Variance}} \\
\textbf{Usefulness}                     & \textit{T}      & \textit{C}      & \textit{T}        & \textit{C}        \\ \hline 
\textit{Question Answering}             & 2.95            & 3.06            & 0.76              & 0.43              \\
\textit{Contextualization of questions} & 2.93            & 2.67            & 0.64              & 0.98              \\
\textit{Statistics}                     & 2.96            & 2.76            & 0.95              & 0.81              \\
\textit{Open feedback}                  & 2.82            & 3.09            & 0.77              & 0.59              \\
\textit{About page}                     & 2.59            & 2.33            & 0.69              & 0.92              \\
\textit{Tokens}                         & 2.71            & 1.88            & 1.07              & 0.80               \\
\textit{Up/ down voting}                & 2.80             & 2.88            & 0.89              & 0.61          \\ 
\textbf{Average}                        & \textit{2.82}   & \textit{2.80\footnote{Because the control group did not utilize tokens, its usefulness evaluation is removed from the calculation of the control groups average.}}   & \textit{0.82}     & \textit{0.73}    \\ \hline
\end{tabular}
\label{tab:usefulness}
\end{table}

Figure \ref{fig:open_feedback_view} illustrates the obtained unsolicited feedback and the participants ranking of these items for the fourth experiment round. A focus group (FG4 in Table \ref{tab:participants}) evaluated the content and the ranking of the unsolicited feedback as useful for improving the library services because the mechanism highlights the most important feedback items. Moreover, the focus group highlighted the innovativeness and useability of combining solicited and unsolicited feedback into one software artifact, because it facilitates the combination of quantitative and qualitative analysis. In particular, the group was positively surprised about the quantity of provided unsolicited feedback, though it was not incentivized. 
Also, the focus group stated that because the existing library wall (see Section \ref{sec:best_practice_mechanisms}) had been utilized in the software artifact, an integration of the tool into the infrastructure and processes of the organization would be simple.

\subsubsection{Feedback Contextualization}
\label{sec:feedback_contextualization}


The participants of the experiment evaluated the contextualization feature of the software artifact as useful (Table \ref{tab:usefulness}). In particular, the contextualization options are neither perceived as restricting nor do users want to have on average more contextualization options (Table 17 of the Supplementary Material). This indicates that the chosen contextualization options are sufficient for the users to express themselves.

The focus group (FG4 in Table \ref{tab:participants}) evaluated the contextualization of solicited and unsolicited feedback as innovative and useful improving the quality of the collected feedback. In particular, the focus group evaluated the contextualization of solicited feedback items as an enabler for i) an identification of weakly formulated questions by analyzing those with low satisfaction rating via the comment contextualization, ii) identification of feedback items that are important for the library to improve their services by focusing on those items that have a high importance rating and iii) a differentiated view on given feedback by comparing rating behavior of all users with those that found the question important to improve the library service (Figure 13 of the Supplementary Material). In particular, this differentiated view on given feedback is described as "very interesting, because it enables a better interpretation of answers". 

The ranking of unsolicited feedback is evaluated as useful because it enables prioritization of unsolicited feedback which is not possible with the current implementation of the feedback wall in the library.
Moreover, the combination of unsolicited with solicited feedback in one software tool has been evaluated "as a very useful approach for the library" as it facilitates a combination of quantitative and qualitative analysis. 

\subsubsection{Token Incentives}
\label{sec:token_incentives}
The token feature of the software artifact is evaluated as useful by the treatment group (Table \ref{tab:usefulness}). In particular, the token carries value for that group (Table \ref{tab:utaut}).
Table \ref{tab:app_interactions} illustrates the impact of token incentives on the amount of provided feedback and contextualizations. On average, the treatment group provided more solicited feedback and contextualizations than the control group. The latter indicates that the incentives are encouraging participants to increase the depth of their feedback and thus its quality \citep{Mudambi_2010, Burtch_20218}. Unaware-users of the library services are incentivized to increase the amount of solicited feedback, indicating the potential of the chosen incentives to mobilize non-customers of the library to provide feedback.  Nevertheless, in the mean, the control group gave more unsolicited feedback than the treatment group, which might be due to the following: Providing unsolicited feedback is not incentivized. Thus, users have to have intrinsic motivation, which might be crowded out in the treatment group with the applied incentives \citep{osterloh2000motivation}.

\section{Discussion}
\label{sec:discussion}

\subsection{Design Principles Revisited}
\label{sec:desgin_principles_revisted}

\begin{table}[tb] \caption{The finalised Design Principles for customer feedback systems per category (C.) illustrating if a principle has been applied in the artifact utilized in the experiment (EX.), how the principle has been implemented in the software artifact (Implementation), if it has been evaluated by the users (U) or focus group (FG), and the findings of these evaluations.} \label{tab:design_principles}
\begin{tabular}{lclclccl} \hline
\textbf{ID} & \multicolumn{1}{l}{\textbf{C.}} & \textbf{Design Principle}                                                                  & \textbf{Ex.} & \textbf{Implementation}                                                                                                                      & \multicolumn{2}{c}{\textbf{Eval.}} & \textbf{Finding}                                                                                                 \\
            & \multicolumn{1}{l}{}            &                                                                                            &              &                                                                                                                                              & U               & FG               &                                                                                                                  \\ \hline
\textit{1}  & \multirow{3}{*}[-1.5em]{\rotatebox[origin=c]{90}{Infrastrucutre}}  & Use a public data store                                                                    & x            & Blockchain (Ethereum)                                                                                                                        &                 &                  &                                                                                                                  \\
\textit{2}  &                                 & \begin{tabular}[c]{@{}l@{}}Use a transparent\\ computing engine\end{tabular}               & x            & Smart Contracts                                                                                                                              &                 &                  &                                                                                                                  \\
\textit{3}  &                                 & \begin{tabular}[c]{@{}l@{}}Create a self-explantory\\ System/UI\end{tabular}               & x            & \begin{tabular}[c]{@{}l@{}}Reused existing interfaces\\ (Google, Reddit, Stackover.)\\ and tools (feedback wall,\\ finance 4.0)\end{tabular} & x               & x                & \begin{tabular}[c]{@{}l@{}}fast onboarding;\\ low-threshould;\\ simple integr. \\ in exist. proces.\end{tabular} \\ \hline
\textit{4}  & \parbox[t]{2mm}{\multirow{4}{*}[-1.5em]{\rotatebox[origin=c]{90}{Feedback}}}        & \begin{tabular}[c]{@{}l@{}}Enable contextualizton\\ of feedback\end{tabular}               & x            & Dedicated views                                                                                                                              & x               & x                & \begin{tabular}[c]{@{}l@{}}exhaustive; impr.\\ post-analysis\end{tabular}                                        \\
\textit{5}  &                                 & Rank feedback                                                                              & x            & \begin{tabular}[c]{@{}l@{}}Up/ down voting of\\ unsolicited feedback\end{tabular}                                                            &                 & x                & \begin{tabular}[c]{@{}l@{}}Enables\\ priorization\end{tabular}                                                   \\
\textit{6}  &                                 & Visualize (feedb.) statistics                                                            & x            & Dedicated statistics view                                                                                                                    & x               &                  & High usefulness                                                                                                  \\
\textit{7}  &                                 & \textit{\begin{tabular}[c]{@{}l@{}}Combine solicited and\\ unsol. feedback\end{tabular}}    & \textit{x}   & \textit{Dedicated views}                                                                                                                     & \textit{}       & \textit{x}       & \textit{\begin{tabular}[c]{@{}l@{}}Quant. and\\ qual. eval\end{tabular}}                                         \\ \hline
\textit{8}  & \parbox[t]{2mm}{\multirow{4}{*}[-2em]{\rotatebox[origin=c]{90}{Interaction}}}   & \begin{tabular}[c]{@{}l@{}}Enable cyber-physical \\ interactions \end{tabular}                                                                  &              & \begin{tabular}[c]{@{}l@{}}Interactive answering\\ options; real-time digital\\ voting centers\end{tabular}                                  &                 &                  &                                                                                                                  \\
\textit{9}  &                                 & \begin{tabular}[c]{@{}l@{}}(Re)use already existing\\ pseudonymous identities\end{tabular} & x            & Blockchain addresses                                                                                                                         &                 &                  &                                                                                                                  \\
\textit{10} &                                 & Build an opt-in system                                                                     & x            & \begin{tabular}[c]{@{}l@{}}No force applied\\ on users to join\end{tabular}                                                                  &                 &                  &                                                                                                                  \\
\textit{11} &                                 & Utilize rewards                                                                            & x            & Blockchain-based tokens                                                                                                                      & x               &                  & \begin{tabular}[c]{@{}l@{}}Improves quant.\\ and quality\end{tabular}      \\ \hline                                       
\end{tabular}
\end{table}

The findings of the previous section illustrates how implementing the identified design principles in a software artefact results in a value-sensitive CFS (answer to RQ 1) that is effective in terms of usability (Section \ref{sec:app_interaction}) and quality of collected feedback (answer to RQ 2). 
Table \ref{tab:design_principles} summarizes them by illustrating these design principles, their implementation in the software artifact, the evaluation of these implementations and the associated findings. In the following the main findings are dicussed:

Due to the Covid-19 policy at the research institute, real-world interactions were restricted. 
Thus, the physicality in the principle of cyber-physical interactions (ID 8 in Table \ref{tab:design_principles}) had not been facilitated in the experiment.  Nevertheless, the value of real-world interactions with humans had been identified as important for the stakeholders in the interviews (Section \ref{sec:identify_problem}) and the design workshops (Section \ref{sec:design_solution}). Thus, we recommend that the impact of mechanisms that require real-world interactions for feedback provision should be evaluated in future work. 

Three of the design principles (ID 1, 2, 8 in Table \ref{tab:design_principles}) are afforded by the blockchain technology which illustrates the useability of this technology for CFS. In particular, blockchain-based cryptoeconomic incentives represent a novel class of incentives that can motivate users to improve the breadth (number of feedback items) and depth (number of contextualization actions) of shared feedback (Section \ref{sec:token_incentives}). Moreover, the utilization of the technology can enhance the trust in the feedback handling process by making it transparent and verifiable.

This work includes the existing unsolicited feedback provision mechanism of the library (library wall, Section \ref{sec:best_practice_mechanisms}) to account for the design principle of self-explanatory system/UI (ID 3 in Table \ref{tab:design_principles}) by reusing interfaces already familiar to the stakeholders. This resulted also in the combination of solicited and unsolicited feedback prevision in one software artifact, which has been identified by the experts of the focus group (FG4 in Table \ref{tab:participants}) as innovative and useful. In particular, it facilitates the combination of quantitative and qualitative analysis. Because of that, we added the principle of combining solicited and unsolicited feedback collection to the list of design principles (ID 7 in Table \ref{tab:design_principles}).

The contextualization of feedback (ID 4 in Table \ref{tab:design_principles}) has been evaluated by both, the experiment participants and the focus group as useful. In particular, amongst others, it enables a differentiated post-analysis of the feedback by comparing rating behavior of users based on the given contextualizations (Section \ref{sec:feedback_contextualization}).

\subsection{Theoretical implications for DSR stemming from a focus on value-sensitive design}
\label{sec:theoretical_implication}

Utilizing value sensitive design in DSR simplifies the design process as it restricts the potential design space of CFS to those configurations that align with stakeholder values. For instance, the design space of blockchain-based systems is large \citep{ballandies2021decrypting}. In particular, one can choose between using a private blockchain (data stored and mechanisms not transparent) or a public blockchain. Each decision coming with different implications for system properties (e.g. level of immutability of data entries, transparency of mechanisms/ smart contracts, etc.) and dependent design options (e.g. the access rights to interact with the system). By utilizing value-sensitive design we obtained the design principles of using a public and immutable storage and a transparent computing engine, which can be facilitated with a permissionless and public blockchain. Thus, the design options associated with private blockchains had been removed upfront and thus simplified/accelerated the design process by providing focus on the solutions using public blockchain infrastructures.

\begin{figure}[tb]
    \centering
    \includegraphics[width=1.0\textwidth]{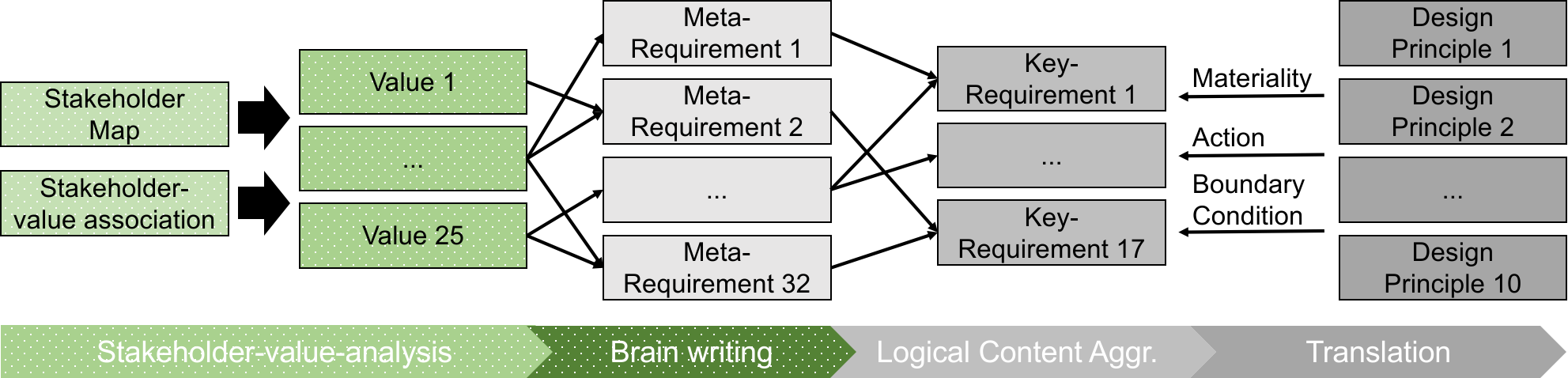}
    \caption{Design principle development according to \citet{moller2020design,chandra2015prescriptive,walls1992building}, as introduced in \citet{hasse2022design}, and extended with value-sensitive design methods (green/ patterned boxes), as introduced in this work: A stakeholder-value analysis (Section \ref{sec:stakeholder_analysis} and \ref{sec:value_analysis}) and brain writing method (Section \ref{sec:design_requirements}) can be utilized to connect meta-requirements of a software artefact with important values that stakeholders carry, eventually resulting in design principles that instantiate those values in an IT artefact.}
    \label{fig:theory}
\end{figure}

Moreover, by first identifying the important values which a system has to incorporate and then associating those values with the design principles, as performed in this work, one explicitly receives the rationale for a design principle in the form of values that are instantiated in a software artefact by following that design principle.
This is illustrated in Figure \ref{fig:design_requirements} for each of the design principles. This supports designers and implementors of a systems by transparently and efficiently communicating ethical implication of an instantiated system in terms of values.
This process of connecting design principles with values is summarized in Figure \ref{fig:theory}.

Finally, considering the importance of values for humans and the findings of this work, the DSR community could start theorizing how value-sensitive design methods can be integrated in established DSR methodologies to make design knowledge comparable across application domains and to enhance the confidence in problem-solution links. For instance, system designers might be able to reuse the design principles associated to values of this work if these values are also required for the software artifact of their application domain. This would eventually reduce construction time as design knowledge could be compared and transferred between application domains via values.

\subsection{Limitations}
Utilizing public blockchains comes with a thread to privacy if sensitive and/or identity revealing information is posted to the blockchain that would then be visible to everyone. Several mechanisms exist to anonymize user input or limit the information reveal. For instance, one way is to only store the hashes of information on-chain and keeping/ storing the data itself off-chain, potentially in a privacy-sensitive way. 
In general, the values of privacy and transparency are often conflicting. One can employ value-sensitive design methods such as to make value conflicts transparent and to resolve them, for instance, via data-sharing coordination~\citep{Pournaras2023}. Though not utilized in this work, Table \ref{tab:values_per_stakeholder} provides a first indication that stakeholders for the CFS considered in this work transparency higher than privacy which resulted in the instantiated system design. In particular, privacy is not considered in the instantiated software artifact, which is communicated transparently in Figure \ref{fig:design_requirements}. In future work one could extend the set of values considered for the CFS (e.g. with privacy), explicitly resolve potential value conflicts (via data-sharing coordination~\citep{Pournaras2023}) and then update the system design accordingly.

Due to the scope defined by the case study (CFS in a library organization), high scalability in terms of user input was not required. Several mechanisms exist to improve the scalability of 2nd layer systems that utilize another blockhain as an infrastructure layer, as it is the case in our instantiated software artefact. One of them is to keep some computations off-chain. In order for the constructed system to be deployed in a global organization with a potential large customer base, its scalability has to be evaluated and be accounted for in its system design, which is left to future work. Moreover, if a high scalibility would be required that cannot be facilitated with the chosen public blockchain (Ethereum, see Section \ref{sec:construct_solution}), utilizing either a permissioned/ private blockchain \citep{ballandies2021decrypting} or newer public blockchains (e.g. Solana \citep{yakovenko2018solana}) might be more suitable. Nevertheless, in this case it would be neccessary to analyse the impact of the choice on the instantiated values in the artifact.

Finally, the integration of value-sensitive design methods into the methodology of an established design science research process facilitated the identification of design principles that explicitly consider stakeholder values. The positive evaluation of the software artifact indicates that this approach is promising to construct systems that are value-sensitive and simultaneously improve the status-quo of a systems functioning. Nevertheless, due to the scope of this research, it has not been explicitly evaluated if the software artifact actually is perceived by the stakeholders as to align with their values, which is left to future work.

\section{Conclusion}
\label{sec:conclusion}

This paper argues that the feedback provision to organizations via software artifacts can be improved by following the design principles identified in this work (Table \ref{tab:design_principles}). In particular, in this way a system can be instantiated that is useable, motivates users to provide feedback and that improves the quality of the collected information. By considering values of stakeholders explicitly in the design steps of an established Design Science Research methodology, this work accounts for both, i) the alignment of the created system with stakeholder values such as credibility and autonomy, and ii) an innovation in the way how feedback is provided to organizations by means of blockchain-based incentivization and contextualized information. 
Hence, the principles (Table \ref{tab:design_principles}) can be utilized by decision makers and managers to create novel value-sensitive and status quo-improving customer feedback systems. 
Moreover, the introduced methodology explicitly provides values as a rationale for design principles and also facilitates the efficient design of software artifacts by reducing the design space of potential system configurations to those that are compatible with stakeholder values.  
Finally, this work shows how blockchain technology enables the three design principles of CFS: public and immutable data storage, transparent computation, and appreciative feedback in the form of token rewards.

The results point to various avenues for future research. Firstly, the software artifact and the instantiated system could be utilized in organizations other than libraries to evaluate the generality of the found design principles and thus increase the confidence in the problem-solution link.
Secondly, the reduced provision of unsolicited feedback by the control group when compared to the treatment group indicates a crowding out of intrinsic motivation. This could be validated in future work by further analyzing the interplay of the applied incentivizes. In particular, because we found that the quality of provided feedback in form of contextualizations is improved by applying cryptoeconomic incentives while increasing the quantity (aligned with other parallel studies~\citep{ballandies2022to,Pournaras2023}), future studies could investigate the impact of varying combinations of cryptoeconomic incentives on the characteristics of provided feedback to identify an optimal combination of incentives.
Third, the DSR community could explore the extent to which design knowledge in the form of design principles can be compared between application domains that require the same values in their design. This would eventually reduce design time as design knowledge could be transferred between application domains via values.


\setstretch{2.0}
\bibliographystyle{apacite}
\typeout{} 
\bibliography{references}

\end{document}